\documentclass[10pt, conference, letterpaper]{IEEEtran}
\IEEEoverridecommandlockouts
\usepackage{cite}
\usepackage{amsmath,amssymb,amsfonts}
\usepackage{algorithmic}
\usepackage{graphicx}
\usepackage{textcomp}
\usepackage{xcolor}
\usepackage{booktabs}
\usepackage{algorithm}
\usepackage{subcaption}
\usepackage{makecell}
\usepackage{tabularx}
\usepackage{booktabs} 
\usepackage{booktabs}
\usepackage{multirow}
\usepackage[normalem]{ulem}
\useunder{\uline}{\ul}{}

\def\BibTeX{{\rm B\kern-.05em{\sc i\kern-.025em b}\kern-.08em
    T\kern-.1667em\lower.7ex\hbox{E}\kern-.125emX}}

\begin{document}

\title{A Geometric Algebra-informed NeRF Framework for Generalizable Wireless Channel Prediction}

\author{\IEEEauthorblockN{Jingzhou Shen$^*$, Luis Lago Enamorado$^*$, Shiwen Mao$^\dagger$, Xuyu Wang\textsuperscript{* \textsection}, }
\IEEEauthorblockA{
$^*$Knight Foundation School of Computing and Information Sciences, Florida International University, Miami, FL 33199, US\\
$^\dagger$Department of Electrical and Computer Engineering, Auburn University, Auburn, AL 36849, US\\
Emails: jshen020@fiu.edu, llago016@fiu.edu, smao@ieee.org, xuywang@fiu.edu}
}

\maketitle
\begingroup\renewcommand\thefootnote{\textsection}
\footnotetext{The corresponding author is Xuyu Wang (xuywang@fiu.edu).}
\endgroup


\begin{abstract}
In this paper, we propose the geometric algebra-informed neural radiance fields (GAI-NeRF), a novel framework for wireless channel prediction that leverages geometric algebra attention mechanisms to capture ray-object interactions in complex propagation environments. Our approach incorporates global token representations, drawing inspiration from transformer architectures in language and vision domains, to aggregate learned spatial-electromagnetic features and enhance scene understanding. We identify limitations in conventional static ray tracing modules that hinder model generalization and address this challenge through a new ray tracing architecture. This design enables effective generalization across diverse wireless scenarios while maintaining computational efficiency. Experimental results demonstrate that GAI-NeRF achieves superior performance in channel prediction tasks by combining geometric algebra principles with neural scene representations, offering a promising direction for next-generation wireless communication systems. Moreover, GAI-NeRF greatly outperforms existing methods across multiple wireless scenarios. To ensure comprehensive assessment, we further evaluate our approach against multiple benchmarks using newly collected real-world indoor datasets tailored for single-scene downstream tasks and generalization testing, confirming its robust performance in unseen environments and establishing its high efficacy for wireless channel prediction.
\end{abstract}


\section{Introduction}
\label{intro}
Modern society has witnessed an unprecedented integration of connected devices into every aspect of daily life. Smart sensors, wearable technology, and autonomous systems form an intricate web of wireless communication that underpins critical infrastructure and personal conveniences. This transformation has elevated the importance of wireless channel modeling to a fundamental challenge in telecommunications. Wireless channel modeling characterizes the propagation of electromagnetic signals through diverse environments, capturing phenomena such as signal attenuation, reflection, diffraction, and scattering~\cite{jiang2020channel, wen2025wrf, wang2025privacy, zhao2024cross}. Precise channel models are vital for effective network design, resource allocation, and quality of service optimization in increasingly dense and heterogeneous wireless networks~\cite{wang2018survey}. 

Traditional modeling approaches, including statistical methods and deterministic ray tracing techniques~\cite{huang2018big, he2018clustering, 7152831}, face limitations in balancing computational efficiency with modeling accuracy, particularly in complex indoor and urban environments. Driven by recent advancements in machine learning, various data-driven approaches are proposed for wireless channel modeling~\cite{yang2019generative,8422221, 5466252, 7792374, 10.1145/3300061.3345438}. However, conventional machine learning models often represent channel characteristics as statistical patterns without incorporating information about the propagation environment. 

The emergence of NeRF~\cite{mildenhall2021nerf} in computer vision provides a novel opportunity to address these challenges. Unlike traditional ray tracing methods that require detailed geometric models and material properties, NeRF can learn signal propagation characteristics directly from measurement data while implicitly capturing environmental geometry. This capability enables accurate channel prediction in environments where obtaining precise 3D models is impractical or costly. Recent work has demonstrated the feasibility of this approach, with studies showing that NeRF variants can effectively capture complex propagation phenomena, including multipath effects and shadowing~\cite{ jia2025neuralreflectancefieldsradiofrequency, shen_mass25}. Furthermore, NeRF's continuous spatial representation allows for channel prediction at arbitrary locations within the modeled space, providing the fine-grained spatial resolution that discrete measurement-based methods cannot achieve. The implicit scene representation learned by NeRF also enables efficient storage and deployment compared to maintaining detailed environmental databases required by deterministic models. Additionally, NeRF's differentiable nature facilitates integration with optimization frameworks for network planning and resource allocation tasks. 

While NeRF demonstrates strong reconstruction capabilities for wireless channel modeling, enabling accurate channel prediction at arbitrary locations through implicit scene representations, it faces several limitations in wireless applications. Originally designed for visual rendering, NeRF does not explicitly model electromagnetic wave phenomena such as reflection coefficients and diffraction patterns, which are fundamental to radio propagation. Additionally, NeRF models exhibit poor generalization to new environments and lack interpretability, preventing researchers from extracting physical insights or validating predictions against electromagnetic theory. While some models address generalization by incorporating spectrum and position information, the inclusion of image data increases both information requirements and training time. These limitations necessitate significant modifications before NeRF can be effectively deployed in practical wireless systems.

\textbf{Challenges.} We face two primary challenges in this work. First, modeling ray-object interactions in wireless environments requires careful consideration of multiple factors. However, limited research has examined how these interactions affect wireless channel modeling. This gap presents substantial obstacles to accurate channel prediction and effective system design. While computer vision research has addressed related problems, for example, NeRFReN~\cite{guo2022nerfren} utilizes separate NeRFs to model direct transmission and reflections, and NeuS~\cite{wang2021neus} incorporates signed distance fields (SDF) to enhance volume rendering, wireless applications introduce unique constraints that differ significantly from vision-based scenarios. Although some studies have adapted these techniques to wireless domains~\cite{orekondy2023winert, 10465179}, current approaches often oversimplify interactions as frequency-independent specular reflections or use opaque neural models with poor generalization. These methods fail to fully exploit NeRF's capabilities for accurate wireless channel modeling. Second, achieving model generalization across diverse environments is critical for practical deployment. Real-world wireless environments vary significantly from training conditions due to differences in building materials, furniture layouts, and structural configurations. These variations directly impact signal propagation patterns. Without robust generalization, models would require retraining for each deployment scenario, making them impractical for network planning across urban and indoor environments. Thus, developing models that maintain accuracy across different physical configurations while adapting to environmental changes is essential for wireless system deployment.

\textbf{Solutions.} To overcome these challenges, we propose a novel multi-view framework that effectively integrates geometric algebra~\cite{dorst2022guided, 10.5555/1610323, roelfs2021gradedsymmetrygroupsplane} with Euclidean algebra. First, we introduce a specialized tokenizer employing multiple algebraic embeddings, allowing the model to learn comprehensive ray-object interactions at both local and global scales. Second, instead of operating on individual sampled points, we utilize a NeRF-inspired approach enhanced with Kolmogorov-Arnold networks (KANs)~\cite{liu2025kan} to generate spatial representations at the ray level, facilitating more robust and physically consistent feature extraction. Third, we propose an attention-driven ray tracing module that leverages learned interactions rather than relying on traditional numerical calculations, significantly improving the accuracy of the final signal strength predictions. Unlike existing NeRF variants that focus on point-to-point interactions, our ray-to-ray paradigm efficiently captures complex spatial dependencies and propagation behaviors that conventional coordinate-based methods fail to represent. Our model demonstrates superior generalization performance compared to recent approaches such as NeRF$^2$~\cite{nerf2} and classic machine learning models in wireless communications. The proposed method achieves improved adaptability across diverse downstream tasks without requiring additional training data. Unlike GWRF~\cite{yang2025gwrfgeneralizablewirelessradiance}, which relies heavily on spectrum image data and incurs significant computational and memory costs, our spatial-data-only approach provides a more efficient and scalable solution for wireless channel modeling.
The main contributions of this paper are summarized as follows:

\begin{itemize}
    \item To the best of our knowledge, this work is the first to leverage cross-algebra representation learning for wireless channel prediction. By encoding signals across multiple geometric spaces, our approach exploits different mathematical structures to capture the complementary aspects of channel behavior. This multi-view strategy enhances the model's ability to identify both structural patterns and invariant properties inherent in wireless channels.
    \item We propose a multi-view tokenizer that enhances embeddings by implicitly learning ray-object interactions while combining Euclidean and geometric algebra advantages.
    \item Our method achieves superior performance in both single-scene evaluation and cross-scenario generalization, outperforming existing baseline approaches. Additionally, we will release a custom-built dataset to support and advance future research in this domain.
\end{itemize}

\section{Preliminary}
\label{sec:Preliminary}
\subsection{NeRF in Wireless Domain}
NeRF$^2$~\cite{nerf2} extends the NeRF framework to model wireless signal propagation. Traditional NeRF methods approximate the scene radiance field by optimizing a neural network $F_{\theta}$ that predicts emitted radiance $\mathbf{c}$ and density $\sigma$ at a 3D location $\mathbf{x}$, given a viewing direction $\mathbf{d}$:
\begin{equation}
(\sigma, \mathbf{c}) = F_{\theta}(\mathbf{x}, \mathbf{d}).
\end{equation}
NeRF$^2$ adapts this formulation specifically for radio-frequency (RF) signals, predicting the attenuation and signal phase components crucial to wireless propagation:
\begin{equation}
(\alpha, \phi) = G_{\psi}(\mathbf{x}, \mathbf{v}, \mathbf{p}_{\text{tx}}),
\end{equation}
where $\alpha$ and $\phi$ represent signal attenuation and phase, respectively, $\mathbf{v}$ denotes the view direction toward the receiver~(RX), and $\mathbf{p}_{\text{tx}}$ is the transmitter position. By incorporating geometric and material-dependent features into neural representations, NeRF$^2$ can capture multipath propagation phenomena, including reflection, diffraction, and scattering. This enables accurate and efficient predictions of complex RF environments, substantially improving wireless channel modeling over traditional methods. Recent studies have explored applying NeRF to wireless field reconstruction~\cite{10.5555/3692070.3693415, yang2025gwrfgeneralizablewirelessradiance, shen2025nerfaptnewnerfframework, yuan2025constructing, tong2025commercial}.

\subsection{Geometric Algebra}
Geometric algebra provides a unified mathematical framework that extends traditional vector algebra by incorporating higher-dimensional geometric objects and operations. This algebraic system, developed by Clifford, enables the representation of rotations, reflections, and other geometric transformations through a single coherent structure~\cite{10.5555/1610323, dorst2022guided, roelfs2021gradedsymmetrygroupsplane}. The geometric algebra $G_{3,0,1}$ represents a four-dimensional space with three spatial dimensions and one temporal dimension, making it particularly suitable for space-time calculations and relativistic physics applications. In this algebra, the fundamental elements include scalars, vectors, bivectors, trivectors, and pseudoscalars, where each element corresponds to different geometric interpretations. The geometric product of two vectors $\mathbf{a}$ and $\mathbf{b}$ is defined as:
\begin{equation}
\mathbf{a}\mathbf{b} = \mathbf{a} \cdot \mathbf{b} + \mathbf{a} \wedge \mathbf{b},
\end{equation}
where $\mathbf{a} \cdot \mathbf{b}$ denotes the symmetric inner product and $\mathbf{a} \wedge \mathbf{b}$ represents the antisymmetric outer product. The algebra $G_{3,0,1}$ operates with a basis $\{1, \mathbf{e}_1, \mathbf{e}_2, \mathbf{e}_3, \mathbf{e}_4\}$ where the spatial basis vectors satisfy $\mathbf{e}_i^2 = 1$ for $i = 1, 2, 3$ and the temporal basis vector satisfies $\mathbf{e}_4^2 = -1$. This structure allows for natural representation of Lorentz transformations and provides computational advantages in physics simulations and computer graphics applications that require efficient handling of rotational and translational operations in space-time. Recent works demonstrate that geometric algebra can be integrated into different machine learning models, resulting in improved performance~\cite{10.5555/3618408.3619627, brehmer2023geometric, zhong2025graph, NEURIPS2024_277628cf}.

\section{Framework}
\label{sec:Framework}

\subsection{Overview}
In this paper, we present GAI-NeRF, a NeRF-based model designed to reconstruct wireless channels using various wireless measurements. Our approach addresses two key objectives: first, we incorporate geometric information into the model to enable learning of ray-object interactions during training; second, we leverage this information to enhance wireless channel modeling with improved accuracy and generalizability. Inspired by classification tokens in large language models (LLMs)~\cite{47751} and multimodal models~\cite{pmlr-v162-li22n}, we develop a multi-view tokenizer that captures both global and local information to inform the learning process.

\begin{figure}[htbp]
\centering
\includegraphics[width=\linewidth]{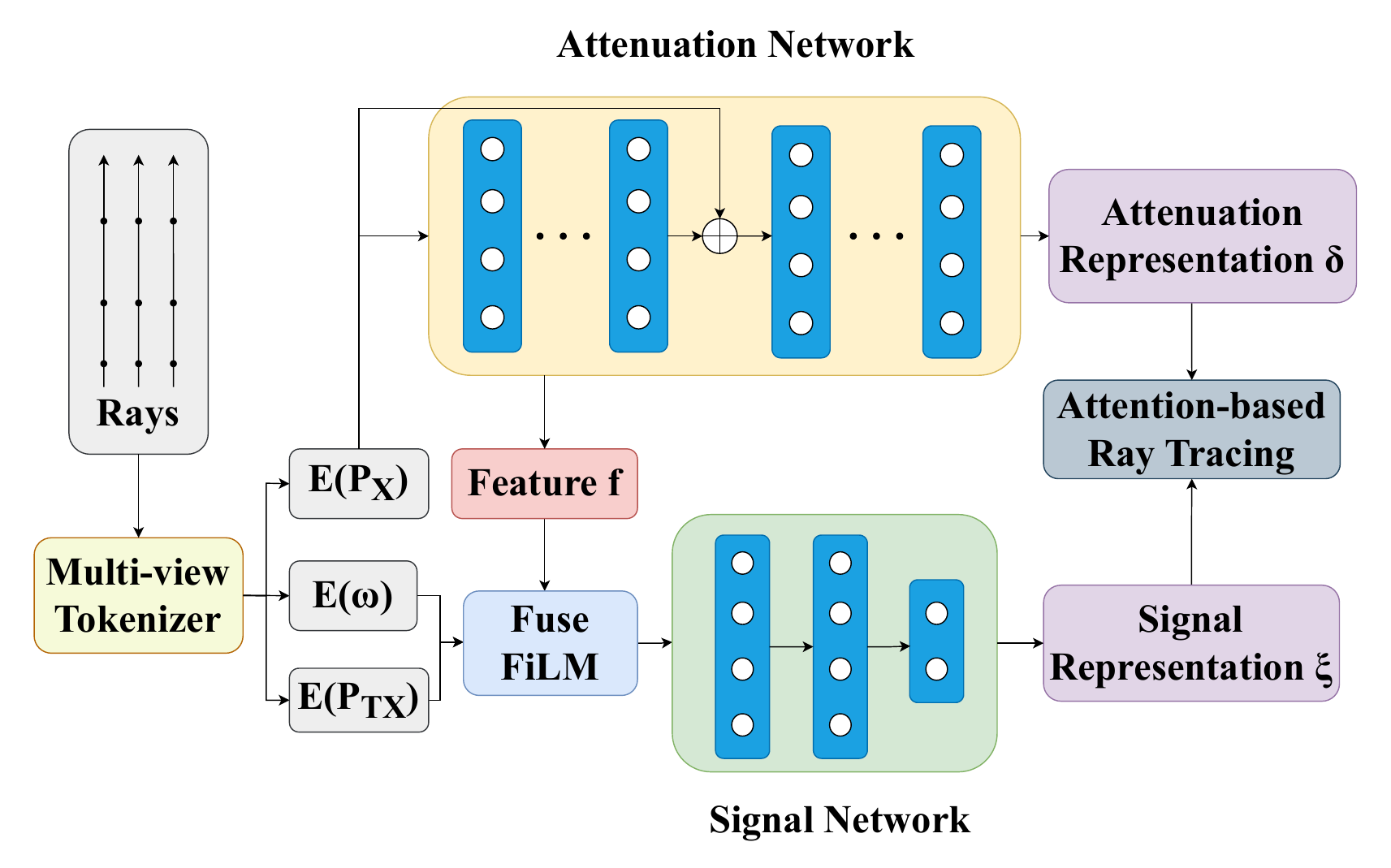}
\caption{Model overview.}
\label{figure_nu}
\end{figure}

\begin{figure}[htbp]
\centering
\includegraphics[width=0.9\linewidth]{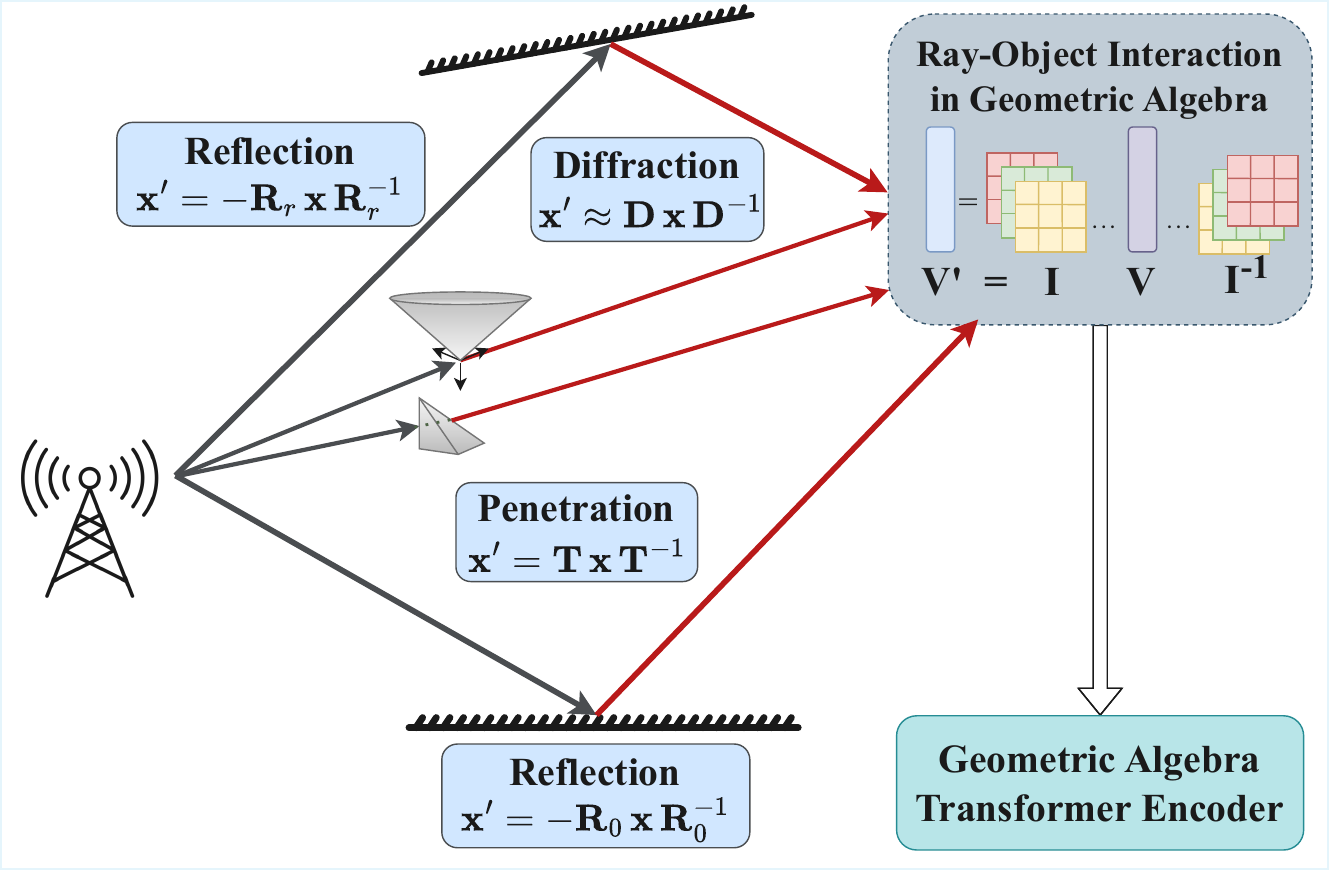}
\caption{Geometric algebra in wireless scene.}
\label{figure_GA}
\end{figure}

Fig.~\ref{figure_nu} illustrates our model architecture, which comprises three main components: a multi-view tokenizer, a NeRF-based model, and an attention-based ray tracing module. Fig.~\ref{figure_GA} illustrates how our tokenizer captures ray-object interactions in wireless environments. The parameters $R_r$, $D$, $T$, and $R_0$ represent rotors for different physical interactions, which are formulated as sandwich products and learned implicitly by our geometric encoder. This unified algebraic representation enables our geometric algebra-based attention mechanism to model complex propagation phenomena within a single mathematical framework. The total transformations in the wireless scene are expressed as $\mathbf{V}' = \boldsymbol{I}\mathbf{V}\boldsymbol{I}^{-1}$, where $\boldsymbol{I}$ represents the combination of multiple ray-object interactions, $\mathbf{V}$ denotes the original ray properties (direction, wave vector, or signal characteristics), and $\mathbf{V}'$ represents the final propagation state after successive interactions including reflection, refraction, and diffraction. This approach eliminates the need for separate specialized modules for each physical effect.

The NeRF-based model includes two subnetworks: the attenuation network and the signal network. The attenuation network processes local embeddings from sampled points along each ray to generate attenuation representations and a feature vector $f$. The signal network takes multiple inputs, including the feature-wise linear modulation~(FiLM) output of feature $f$, local embeddings of view directions, transmitter positions, and global embeddings of these parameters and sample points. This network produces signal representations that undergo mean pooling before concatenation. The concatenated representations then pass through the attention-based ray tracing module to generate the final outputs.

\subsection{Multi-view Tokenizer}
\begin{figure}[htbp]
\centering
\includegraphics[width=\linewidth]{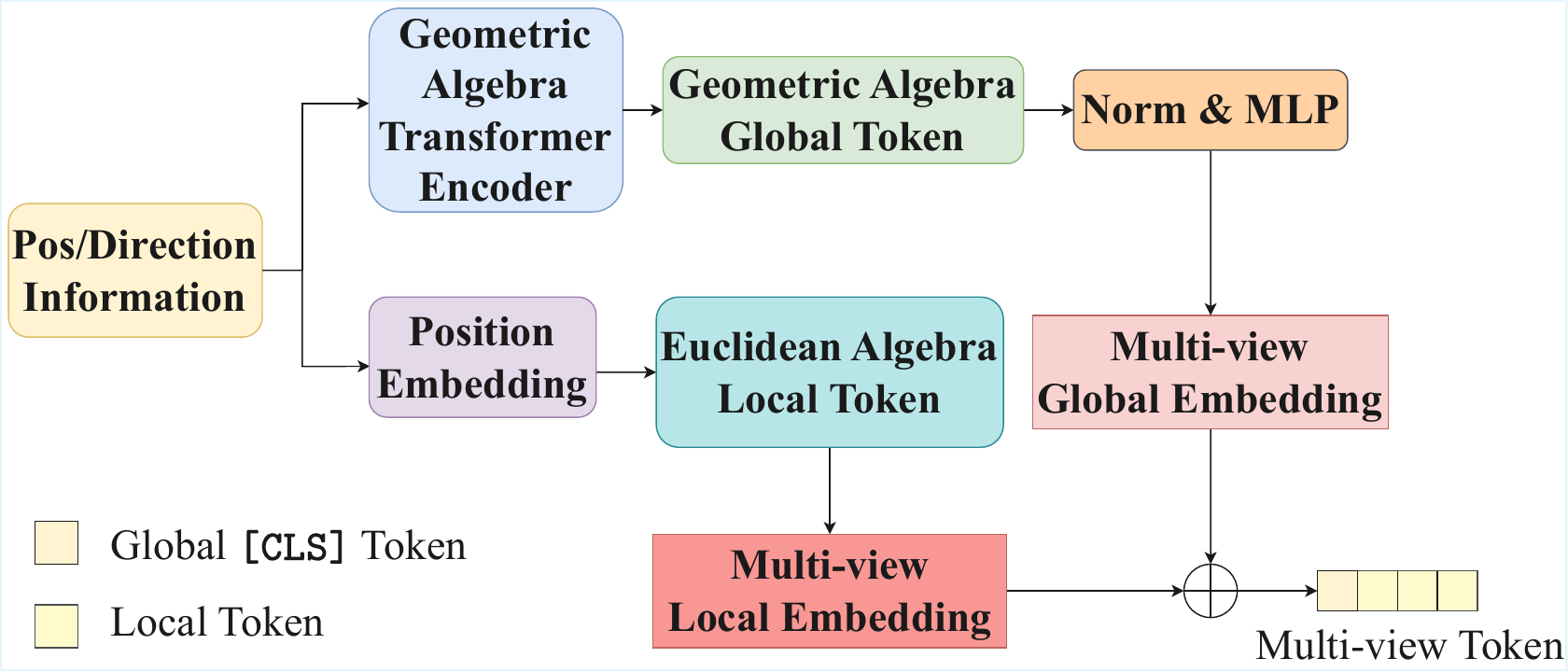}
\caption{Multi-view tokenizer overview.}
\label{figure_tokenizer}
\end{figure}

We next elaborate on how to design the multi-view tokenizer, shown in Fig.~\ref{figure_tokenizer}, to obtain the local and global embeddings for the entire wireless scenes.

\subsubsection{Geometric Algebra Transformer Encoder} 
We employ the geometric algebra transformer (GATr)~\cite{brehmer2023geometric} as an encoder to extract initial global embeddings before feeding data into our model. In geometric algebra G(3,0,1), transformations such as rotations, reflections, and diffractions are elegantly represented through sandwich products. This mathematical framework provides a unified approach to describe ray-object interactions in wireless communications. The attention mechanism in neural networks can also be formulated as a sandwich product, enabling us to use geometric attention to implicitly learn spatial ray-object interactions. This structure preserves vector grade and maintains algebraic consistency across different wireless propagation scenarios.

In wireless ray propagation, electromagnetic waves interact with surfaces through sandwich product representations. When a ray ${x}$ encounters surface normal $\hat{n}$, reflection follows the transformation: ${x}' = -{R}{x}{R}^{-1},$ where $R$ corresponds to $R_r$ and $R_0$ in Fig.~\ref{figure_GA}.  For edge diffraction, the transformation employs diffraction operator $D$: $\mathbf{x}' \approx D\mathbf{x} D^{-1}, $ where $D$ can only approximate the physical process, since diffraction alters amplitude, phase, and spatial distribution in ways that cannot be captured by a strict algebraic mapping. Penetration through materials can be represented similarly through sandwich products with appropriate material operators. Multiple interactions along ray paths compose as sequential sandwich product operations:
\begin{equation}
\begin{split}
V' &= I_1 I_2 \cdots I_n V I_n^{-1} \cdots I_2^{-1} I_1^{-1} \\
&= I V I^{-1},
\end{split}
\end{equation}
where each $I_i$ represents an individual ray-object interaction.

The geometric algebra attention mechanism exhibits structural correspondence with physical transformations. We derive the attention operation from input tensors $q$, $k$, and $v$, which contain $n_c$ channels. The standard attention operation:
\begin{equation}
\text{Attention}(q, k, v)_{i'c'} = \sum_i \text{Softmax}_i \left( \frac{\langle q_{i'c}, k_{ic} \rangle}{\sqrt{8n_c}} \right) v_{ic'}
\end{equation}
can be approximately transformed to the sandwich product form: $\text{Attention}(q, k, v)_{i'c'} = \sum_i{A}_{i'}v_{ic'}A_{i'}^{-1}$, where $A_{i'}$ represents the geometric operator from attention weights, $i, i'$ refer to label indices, and $c, c'$ correspond to channel indices.

The geometric algebra framework provides structural advantages by enabling early encoding of scene geometry and electromagnetic interactions. These identical sandwich product formulations directly map spatial and wave propagation characteristics to physically meaningful operations. This approach reduces computational complexity while naturally capturing fundamental electromagnetic principles within a unified mathematical structure. The framework captures reflection, diffraction, and penetration effects through implicit learning of transformations that model electromagnetic ray-object interactions in wireless environments.

Moreover, this approach markedly enhances model generalizability by embedding physical laws directly into the learned representations. Geometric algebra inherently preserves rotational and reflectional symmetries, allowing the model to generalize across different orientations and environmental configurations without extensive retraining. Unlike traditional statistical methods, the physics-based nature of the sandwich product ensures the model relies on universal wave propagation patterns rather than environment-specific correlations. The model therefore maintains accuracy in previously unseen wireless scenarios with minimal additional data or adaptation.

To effectively utilize these learned ray-object interactions, we extract the encoded information and generate multi-view global \texttt{[CLS]} tokens. These embeddings are integrated into our model through a mechanism similar to the  token approach in LLMs. This integration enables the model to leverage the comprehensive understanding of electromagnetic interactions, leading to improved performance and enhanced generalizability across diverse wireless environments. The global \texttt{[CLS]} token serves as a universal token that aggregates information from the entire input sequence. In standard transformers~\cite{vaswani2017attention}, the final encoding of the \texttt{[CLS]} token captures a global representation of the input through attention mechanisms. Its output encoding is used for sequence-level tasks because it attends to all other tokens and extracts the most relevant features for downstream applications.

When employing a GATr encoder to acquire \texttt{[CLS]} token, we gain distinct advantages for wireless communication scenarios. This geometric \texttt{[CLS]} token learns to encode the collective ray-object interactions across the entire scene. Our GATr encoder provides rotation and reflection equivariance for the global \texttt{[CLS]} token, enabling natural adaptation to different wireless environment orientations. The token develops a complete geometric understanding of wave propagation patterns within a single multivector representation that directly captures electromagnetic interaction structures.

\subsubsection{Local and Global Embeddings}
Our geometric algebra global token, obtained from GATr encoder, passes into the Norm \& MLP module to enhance representation quality without information compression. The multi-layer perceptron~(MLP) functions as a feature refinement module that implicitly distills ray–object interaction information from the GATr output, providing critical guidance for subsequent NeRF-variant learning. This refinement not only enhances the quality of the representations but also maintains the essential geometric information encoded by GATr, thereby enabling the model to effectively capture the underlying electromagnetic propagation dynamics and improving the accuracy of wireless channel modeling. Then, we use standard positional embeddings to generate the Euclidean algebra local token. After processing, the global embedding represents scene-level electromagnetic interactions, while local embedding preserves spatial relationships within the propagation context. This framework captures both channel-level characteristics and scene-specific signal behaviors for wireless channel modeling. Finally, embeddings are concatenated as the multi-view token and then fed into the following NeRF-based model, integrating geometric interactions with wave patterns. Specifically, we regard the multi-view global embedding of the concatenated token as the global \texttt{[CLS]} token.



\begin{figure}[htbp]
\centering
\includegraphics[width=0.9\linewidth]{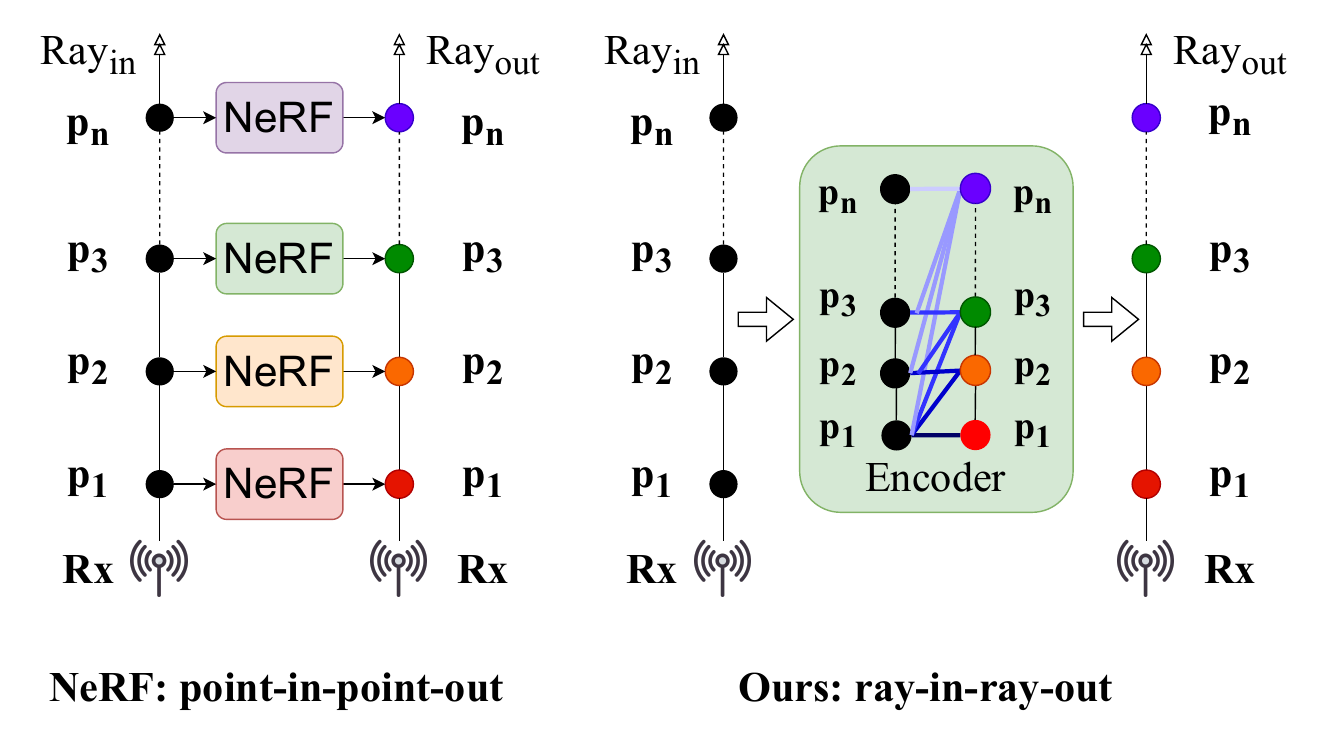}
\caption{Comparison of our ray-in-ray-out and point-in-point-out models with backtraced propagation paths from Rx.}
\label{figure_p2pr2r}
\end{figure}

\subsection{GAI-NeRF Model}
In this subsection, we introduce how to structure our GAI-NeRF to obtain representations of attenuation of signal.
\subsubsection{Ray-to-Ray Model}
We adopt a ray-in-ray-out framework~\cite{wang2022next} in lieu of the traditional point-in-point-out approach. As illustrated in Fig.~\ref{figure_p2pr2r}, in contrast to existing NeRF variants that utilize spatial points on the ray as encoder inputs, our method processes entire rays as the fundamental input units to the NeRF encoder. The ray-based representation captures directional propagation characteristics and spatial coherence of electromagnetic waves while encoding angular information essential for multipath modeling. Our approach enables efficient computation of path loss and phase variations by modeling continuous wave propagation along entire ray paths rather than at isolated spatial points, thereby ensuring closer alignment with the physical principles of wireless signal propagation.

\subsubsection{Model Structure}
Following the NeRF$^2$ framework, our architecture consists of two distinct subnetworks: the attenuation network and the signal network. We replace standard MLPs with efficient KANs using PowerMLP~\cite{Qiu_Miao_Wang_Zhu_Yu_Gao_2025}. PowerMLP is an efficient variant of the KANs, designed to enhance computational speed while preserving interpretability and performance. Unlike standard KANs, which use spline functions, PowerMLP employs a simpler polynomial-like activation known as the rectified power unit~(RePU). Specifically, the RePU activation function is defined as:
\begin{equation}
\text{RePU}(x) = (\text{max}(0, x))^p,
\end{equation}
where $p$ is a learnable exponent parameter. By dynamically adjusting the exponent during training, PowerMLP captures nonlinear relationships without requiring complex spline computations, thus reducing computational overhead.

The attenuation network models how signal power decays as it propagates through the wireless environment. It predicts the path-loss coefficient at each spatial location, determining how much energy is absorbed or scattered along each ray. This network directly influences coverage and channel reconstruction accuracy by controlling spatial signal attenuation patterns. It consists of eight layers of PowerMLPs and outputs the attenuation representation $\delta$. This representation is highly related to the materials of the voxel and independent of the incoming signals. Also, we implement skip-connection in the middle of those layers to allow gradients and features to bypass intermediate layers, facilitating training and preserving essential information across depths.

The signal network handles the actual radiance computation and color prediction. This component processes the attenuated wireless information to generate the final pixel colors observed from different viewpoints. The signal network captures the directional properties of wireless channel propagation and scattering characteristics, enabling accurate channel state reconstruction across various angular distributions. The structure of signal network is similar to the attenuation network. We also use PowerMLPs to replace, and generate the signal emission representation $\xi$. The signal network processes three key inputs: the attenuation feature \textit{f}, embeddings of transmitter positions, and view direction encodings. Before feeding these inputs into the subnetwork, we apply the FiLM module to capture spatial correlations and view-dependent interactions efficiently. FiLM is a conditioning mechanism that modulates neural network activations through learned affine transformations, enabling effective integration of external information by applying element-wise scaling and shifting operations to feature maps based on conditioning inputs. The Performer module enables linear-complexity attention computation while allowing the network to focus on relevant spatial locations. This mechanism helps the signal network identify critical geometric features and establish relationships between different positions within the propagation environment, thereby enhancing the network’s ability to model complex electromagnetic interactions and location-dependent channel effects.


\subsubsection{Attention-based ray tracing Module}
\begin{figure}[htbp]
\centering
\begin{subfigure}{0.52\linewidth}
    \centering
    \includegraphics[width=\linewidth]{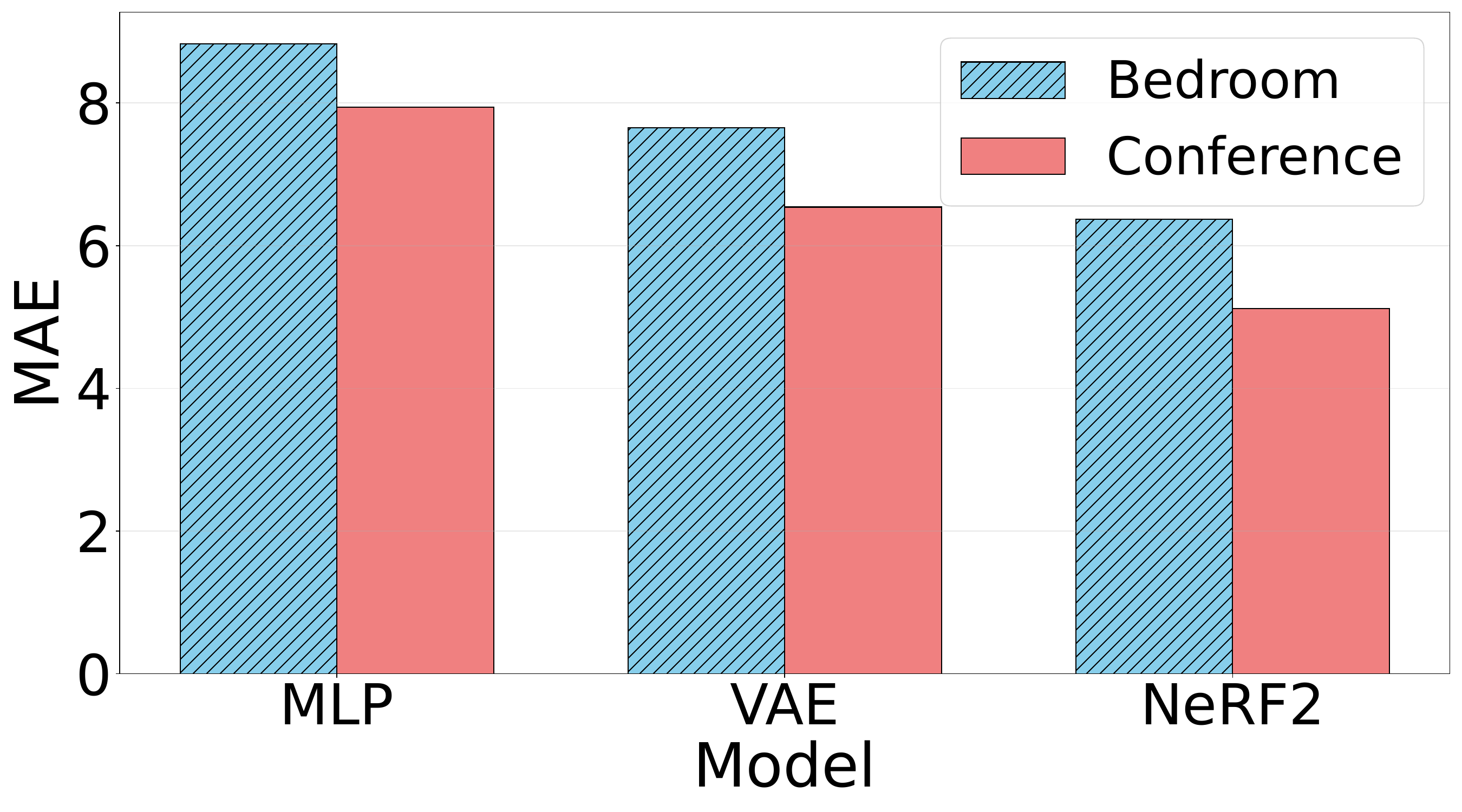}
    \caption{Pilot generalization results.}
    \label{g1}
\end{subfigure}
\hfill
\begin{subfigure}{0.46\linewidth}
    \centering
    \includegraphics[width=\linewidth]{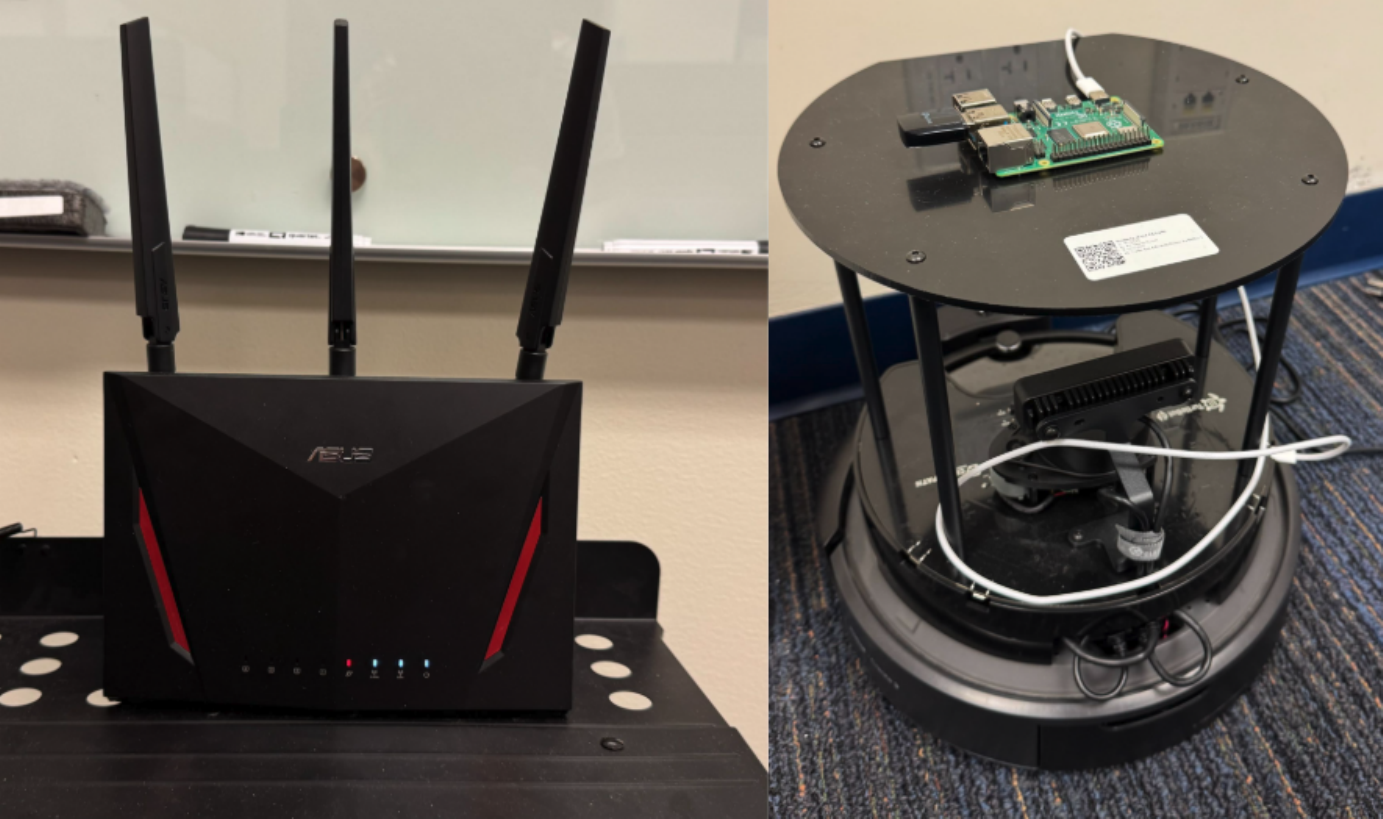}
    \caption{Data collection devices.}
    \label{fig:scene}
\end{subfigure}
\caption{Exploratory performance evaluation and experimental setup.}
\label{fig:combined}
\end{figure}

As illustrated in Fig.~\ref{g1}, we first conduct exploratory experiments to evaluate generalization performance on synthetic datasets~\cite{10.5555/3692070.3693415}. We utilize datasets containing two initial scenes, training our models on one dataset and evaluating on the other one, with mean absolute error~(MAE) serving as the primary validation metric. The results show that basic machine learning models achieve better performance than more complex alternatives. This improved generalization may be attributed to the fact that statistical calculations used in rendering can limit the model's ability to generalize across different datasets.



To improve generalizability across diverse scenarios, we approximate the ray tracing module through attention-based mechanisms. The volumetric ray tracing equation combines attenuation and signal emission through an integral formulation that naturally aligns with attention mechanisms. The standard ray tracing equation computes the observed radiance $C(\mathbf{r})$ along a ray as:
\begin{equation}
C(\mathbf{r}) = \int_{t_n}^{t_f} T(t) \cdot \sigma(t) \cdot c(t) \, dt,
\end{equation}
where $T(t) = \exp\left(-\int_{t_n}^{t} \sigma(s) \, ds\right)$ represents the transmittance function, $\sigma(t)$ is the density coefficient, and $c(t)$ is the emitted radiance at position $t$ along the ray. The transmittance term $T(t)$ acts as an exponential weighting function that determines how much electromagnetic radiation from each point reaches the receiver. This weighting mechanism mirrors the attention mechanism in neural networks, where exponential functions create soft attention weights that emphasize relevant spatial locations. The discretized form of this continuous equation can be represented as follows:
\begin{equation}
C(\mathbf{r}) = \sum_{i=1}^{N} w_i \cdot c_i,
\end{equation}
where $w_i = T_i \cdot \sigma_i \cdot \Delta t_i$ represents the attention weights and $c_i$ is the color value at sample point $i$. The similarity between ray tracing weights and attention weights enables neural networks to learn the volumetric rendering process through attention mechanisms, allowing the network to focus on important spatial regions while maintaining the physical interpretation of electromagnetic wave propagation.

\begin{figure}[htbp]
\centering
\includegraphics[width=0.9\linewidth]{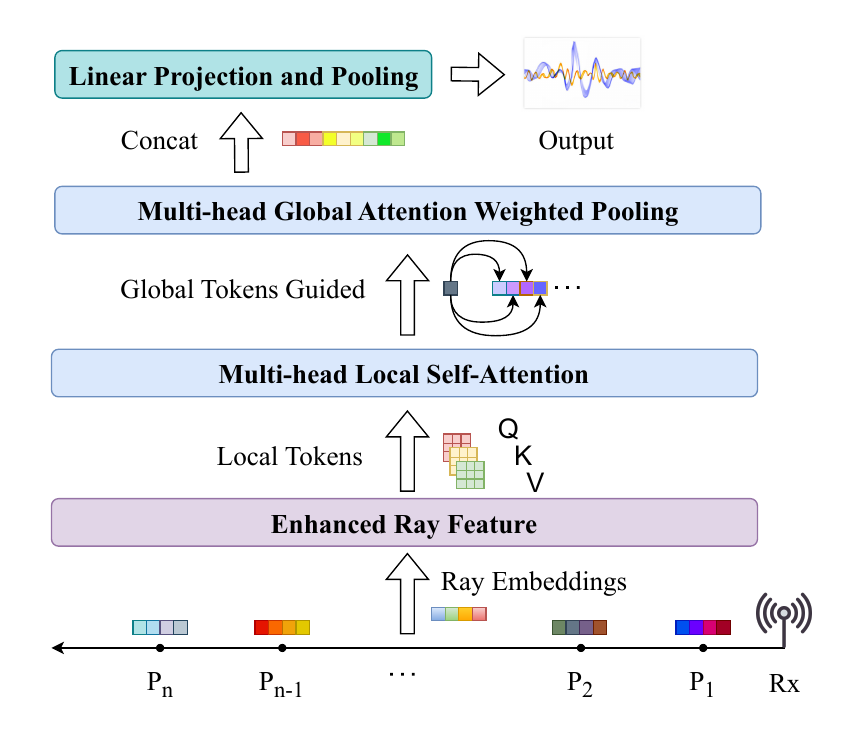}
\caption{Attention-based ray tracing module.}
\label{figure_art}
\end{figure}

Fig.~\ref{figure_art} illustrates our attention-based ray tracing module architecture. After obtaining the outputs, where the global \texttt{[CLS]} token enhances scene representation and improves the accuracy and generalization of our GAI-NeRF, we generate ray embeddings from the rays and their sampled points. These embeddings are added to the input to produce enhanced ray features. We then separate the global \texttt{[CLS]} token from the local tokens. The local tokens pass through a multi-head local self-attention module to obtain initial results. Similar to the GWRF framework~\cite{yang2025gwrfgeneralizablewirelessradiance}, we use ten attention heads corresponding to all possible pairwise interactions between field components, where each attention head captures specific interaction patterns between real and imaginary parts of complex field representations. 
To reduce memory consumption and accelerate training, we employ Performer~\cite{choromanski2021rethinking} as our attention mechanism. Performer is a linear attention mechanism that approximates the standard softmax attention in Transformers using random feature maps. The key innovation lies in decomposing the attention matrix $A_{ij} = \exp(q_i^T k_j / \sqrt{d})$ into a product of two smaller matrices through the kernel trick. Specifically, Performer uses positive random features $\phi(x)$ to approximate the exponential kernel. This allows the attention computation to be reformulated as:

\begin{equation}
\begin{aligned}
\mathrm{Attention}(Q, K, V) 
&= \mathrm{softmax}\!\Bigl(\frac{QK^\top}{\sqrt{d}}\Bigr)\,V \\[6pt]
&\approx \;\phi(Q)\,\bigl(\phi(K)^\top V\bigr).
\end{aligned}
\end{equation}

After obtaining voxel tokens from local self-attention, we calculate attention weights for each ray guided by the \texttt{[CLS]} token, then pool each ray based on these learned weights. These weights indicate how obstacles affect ray propagation paths using the ray-object interaction knowledge encoded in the \texttt{[CLS]} token. Finally, the weighted outputs are concatenated with the global \texttt{[CLS]} token and then passed through a linear projection layer to produce the final predictions.

Our ray tracing approach offers significant advantages over the original method by using a learnable \texttt{[CLS]} token to aggregate global scene information. This token works alongside local tokens to create a centralized representation that summarizes ray-object interactions across the entire wireless scene. Compared to the original GWRF ray tracing implementation, our design significantly improves modeling flexibility as well as generalization and predictive performance.


%

  

\section{Experimental Evaluation And Analysis}
\subsection{Our Own Datasets}

\subsubsection{Experiment setup}
We developed a semi-automated platform for comprehensive indoor received signal strength indicator (RSSI) mapping across the two frequency bands of 2.4 GHz and 5 GHz. The experimental setup employed a human-operated robotic system, shown in Fig.~\ref{fig:scene}, to collect the RSSI measurements throughout a 35 $\textrm{m}^2$ indoor environment over an 18-day period. While the robot positioning throughout the indoor environment was manually controlled, the measurement procedures were fully automated to ensure consistency across measurements and prevent human direct intervention from affecting the signal propagation.

\subsection{Hardware Configuration}\label{sec:hard_config}


The mobile measurement platform, illustrated in Fig.~\ref{fig:scene}, consists of a TurtleBot 4 running on Ubuntu 24.04 with ROS2 Jazzy. The TurtleBot 4 integrates an RPLIDAR A1M8 sensor, which provides $360^\circ$ laser scanning capabilities with a 12 m range, enabling robust Simultaneous Localization and Mapping (SLAM) functionalities. The platform utilizes the Nav2 navigation stack with Adaptive Monte Carlo Localization (AMCL) for position estimation. The SLAM system enables real-time 2D occupancy grid generation while simultaneously tracking the robot's pose within the generated map. This enables precise ($x$, $y$, $z$) coordinate acquisition, which is required for spatially-referenced RSSI measurements.  

\begin{figure}[htbp]
    \centering
    \begin{subfigure}[t]{\linewidth}
        \centering
        \includegraphics[width=\linewidth]{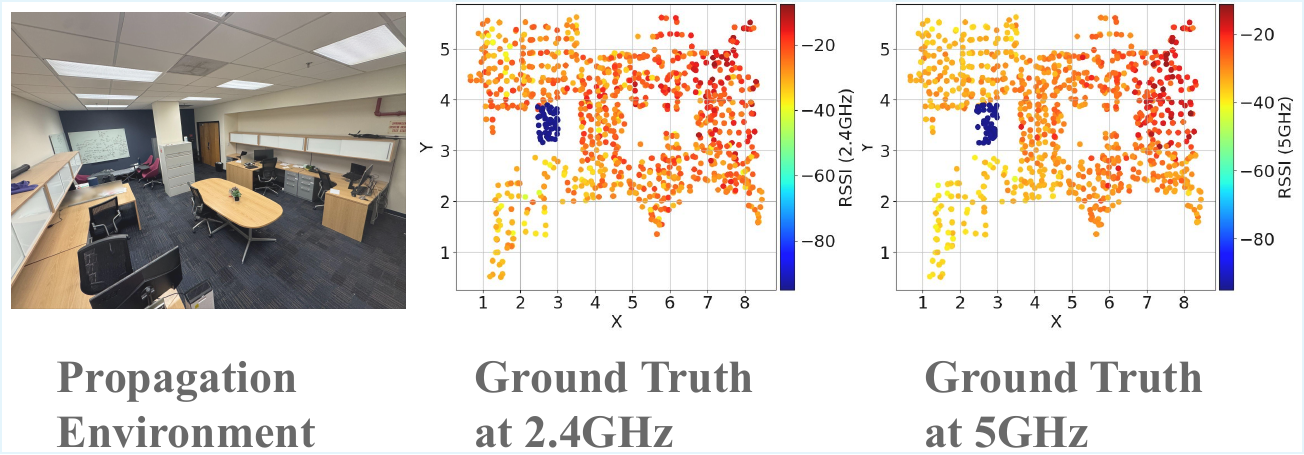}
        \caption{Room 1 setting and ground truth.}
        \label{fig:room1}
    \end{subfigure}
    
    \vspace{0.1cm} 
    
    \begin{subfigure}[t]{\linewidth}
        \centering
        \includegraphics[width=\linewidth]{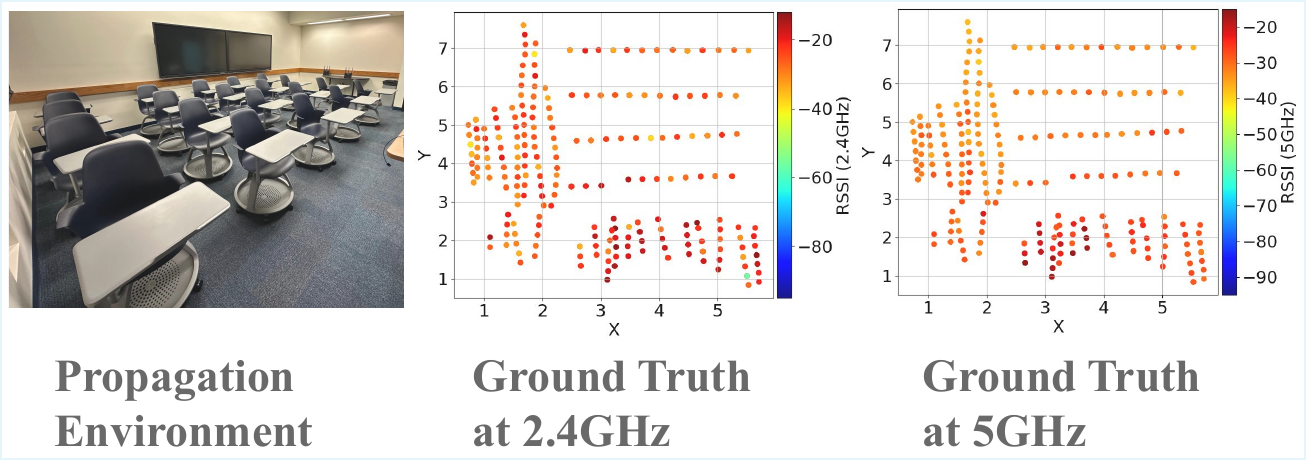}
        \caption{Room 2 setting and ground truth.}
        \label{fig:room2}
    \end{subfigure}
    \caption{Propagation environments and ground truth of our dataset.}
    \label{fig:four-wide}
\end{figure}

A Raspberry Pi 4 Model B (4 GB of memory) served as the primary measurement unit, mounted and fixed at the geometric center of the TurtleBot 4's upper platform. To support RSSI measurements, a TP-Link Archer T3U AC1300 Mini MU-MIMO USB 3.0 adapter, based on a Realtek RTL8812BU chipset, is connected to the Raspberry Pi. The adapter supports dual-band operation (2.4 GHz and 5 GHz) with $2\times 2$ spatial streams. The measurement infrastructure was supported by two ASUS RT-AC86U Dual-Band Wireless-AC2900 routers. 

The data collection procedure combined manual navigation of the TurtleBot 4 with automated RSSI measurements. Using a joystick control, the TurtleBot 4 was manually navigated to a predefined location on the map, after which the control system automatically recorded the robot's coordinates via the \texttt{/amcl\_pose} ROS 2. Subsequently, the control virtual machine (VM) initiated a secure shell (SSH) connection to the Raspberry Pi and executed a script that implemented five consecutive RSSI measurements with a 1 s interval for both frequency bands. Each final measurement was considered valid only if all five samples were successfully collected. Incomplete measurements were automatically discarded and repeated at the same location until five samples were successfully recorded. 
Fig.~\ref{fig:four-wide} displays the rooms and ground truth in our two experimental scenes. The blue regions represent columns, which provide clear obstructions for detecting ray-object interactions within the indoor environment.


\subsection{CSI Datasets from NeRF$^2$ and NewRF}
We further use a few more datasets mentioned in NeRF$^2$ and NewRF~\cite{10.5555/3692070.3693415}. From NeRF$^2$, we used a MIMO-CSI dataset, i.e., the Argos channel dataset~\cite{shepard2016understanding}, which contains a total of 100 K measurements. On the other hand, we use the simulated data from the NewRF repository~\cite{10.5555/3692070.3693415}. Each simulated environment features a single transmitter with fixed placement and 443,975 receiver locations distributed throughout the three-dimensional space. The channel measurements incorporate a 52-subcarrier orthogonal frequency-division multiplexing~(OFDM) configuration to capture frequency-selective fading characteristics.

We address signal distortions in channel estimation by applying a three-step standardization process in FIRE~\cite{10.1145/3447993.3483275}. These transformations remove uniform distortions, including global phase offsets and linear phase trends, while preserving key channel features such as inter-antenna amplitude/phase relationships and the non-linear frequency response that characterizes the wireless channel.

\subsection{Baselines}
Our approach is compared with four distinct baseline architectures. The first baseline employs the MLP structure containing four layers with respective hidden dimensions of 256, 128, 64, and 64 units, culminating in a linear classification layer. The second baseline incorporates a variational autoencoder~(VAE) framework derived from FIRE~\cite{10.1145/3447993.3483275}, featuring three encoding layers paired with four decoding layers that maintain equivalent hidden unit specifications. The third comparative model utilizes a multi-layered perceptron implementation of DCGAN~\cite{radford2015unsupervised}, constructed with four layers each for both generator and discriminator components, where the initial hidden dimension of 512 units systematically reduces by fifty percent across successive layers. The fourth baseline follows the established NeRF$^2$ architectural framework as specified in their original implementation. For the final baseline evaluation, we adopt the ray tracing module from GWRF~\cite{yang2025gwrfgeneralizablewirelessradiance} to assess generalizability. However, GWRF demonstrates poor performance with RSSI data, generating uniform predicted values across the output matrix. This limitation occurs because our RSSI data exhibits significantly lower variance than the public CSI data, and the sum and mean pooling in GWRF compresses data variance, further preventing the model from detecting subtle differences in low-variance datasets. NewRF~\cite{10.5555/3692070.3693415} faces similar challenges as its sampling mechanism produces extremely small or zero weights in our scenarios, which discards important information in our datasets. These fundamental design limitations led to the exclusion of both models from some evaluations. We evaluate performance using the median of MAE and signal-to-noise ratio~(SNR) as our primary metrics.

\begin{table*}[t]
    \centering
    \caption{Performance comparison of GAI-NeRF against baseline methods across different room configurations and CSI scenarios. Lower MAE values and higher SNR values indicate better performance.}
    \renewcommand{\arraystretch}{1.1}
    \setlength{\tabcolsep}{3pt}
    \footnotesize
    \begin{tabular*}{\textwidth}{p{1.8cm}@{\extracolsep{\fill}}|cc|cc|ccc}
        \toprule
        \multirow{3}{*}{\textbf{Method}} & 
        \multicolumn{2}{c|}{\textbf{Room 1 (Our RSSI Dataset)}} & 
        \multicolumn{2}{c|}{\textbf{Room 2 (Our RSSI Dataset)}} &
        \multicolumn{3}{c}{\textbf{Other CSI Datasets}} \\
        \cmidrule(lr){2-3} \cmidrule(lr){4-5} \cmidrule(lr){6-8}
        & \multicolumn{4}{c|}{\textbf{MAE (dB) $\downarrow$}} & \multicolumn{3}{c}{\textbf{SNR (dB) $\uparrow$}} \\
        \cmidrule(lr){2-5} \cmidrule(lr){6-8}
        & \textbf{2.4 GHz} & \textbf{5.0 GHz} & \textbf{2.4 GHz} & \textbf{5.0 GHz} & \textbf{MIMO} & \textbf{Bedroom} & \textbf{Conference} \\
        \midrule
        MLP           & 7.31 & 9.28 & 8.15 & 9.87 & 19.86 & 17.67 & 16.12 \\
        VAE           & 5.75 & 5.49 & 4.53 & 2.73 & 21.47 & 15.11 & 18.90 \\
        DCGAN         & 4.01 & 3.39 & 4.22 & 2.95 & 13.12 & 14.10 & 13.45 \\
        NeRF$^2$      & 3.58 & 2.87 & 3.92 & 2.07 & 23.85 & 21.65 & 20.88 \\
        GWRF          & --- & --- & --- & --- & 25.08 & 28.35  & 26.92 \\
        \midrule[\heavyrulewidth]
        \textbf{GAI-NeRF} & \textbf{3.17} & \textbf{1.75} & \textbf{3.06} & \textbf{1.80} & \textbf{25.78} & \textbf{29.13} & \textbf{27.54} \\
        \bottomrule
    \end{tabular*}
    \label{tab:performance_comparison}
\end{table*}

\subsection{Evaluations}
\subsubsection{Single-scene performance}

\begin{figure*}[htbp]
  \centering
  \newlength{\threecolwidth}
  \setlength{\threecolwidth}{\dimexpr(\textwidth-1.0\columnsep)/3\relax}
  
  \begin{subfigure}[t]{\threecolwidth}
    \centering
    \includegraphics[width=1\linewidth, height=6.5cm, keepaspectratio]{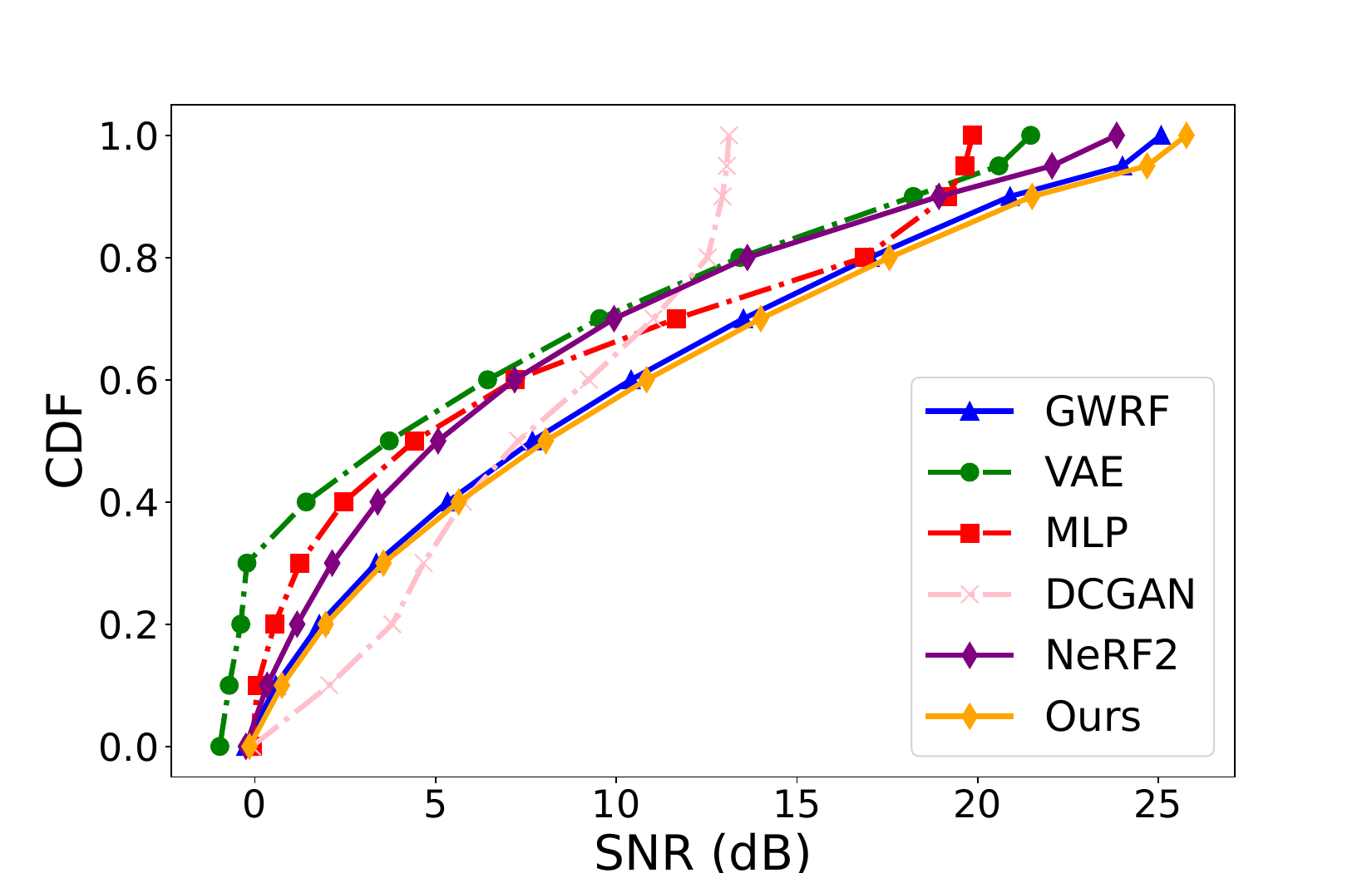}
    \caption{MIMO-CSI dataset.}
    \label{fig:sub1}
  \end{subfigure}\hfill
  \begin{subfigure}[t]{\threecolwidth}
    \centering
    \includegraphics[width=1\linewidth, height=6.5cm, keepaspectratio]{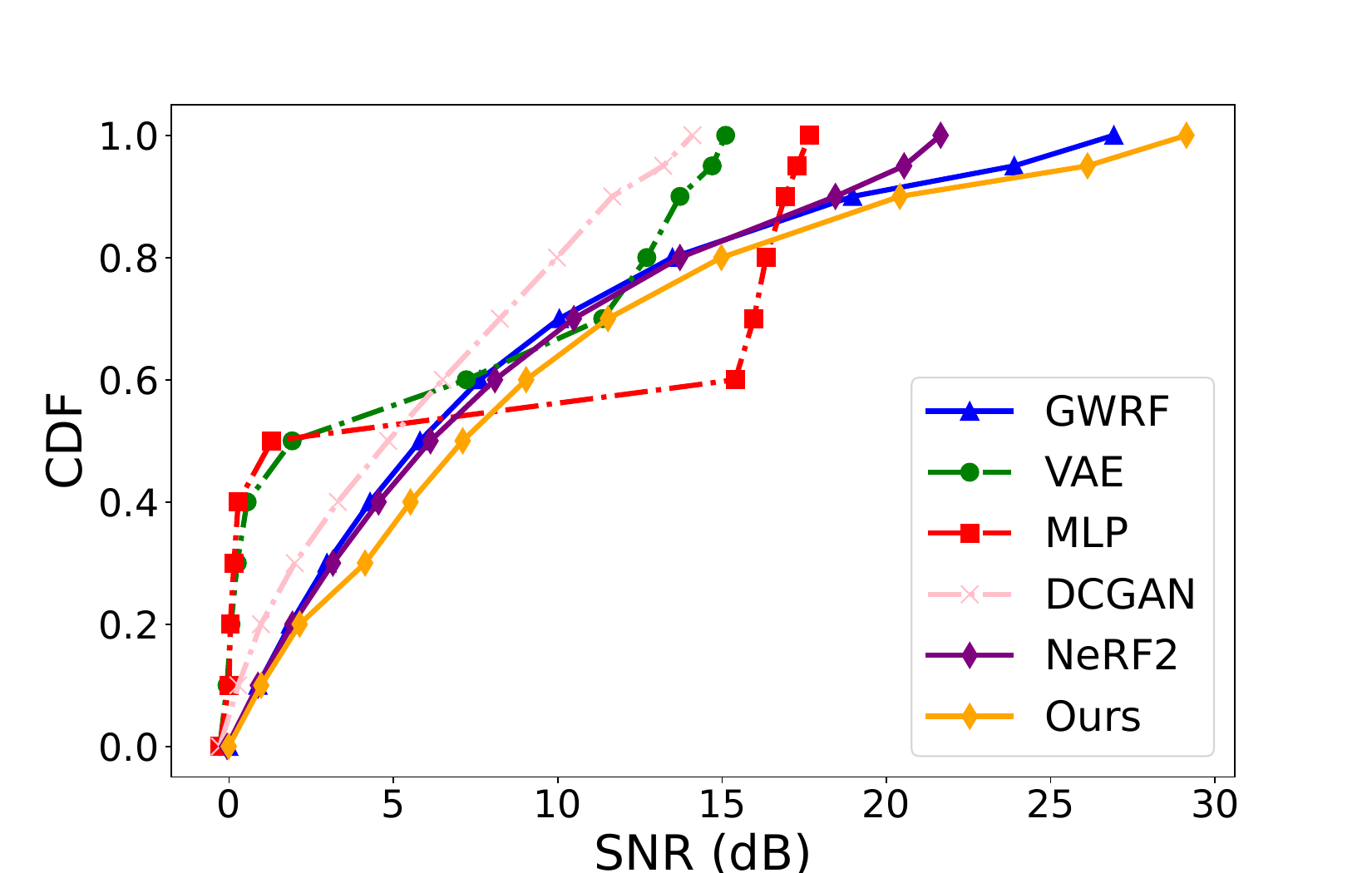}
    \caption{Bedroom dataset.}
    \label{fig:sub2}
  \end{subfigure}\hfill
  \begin{subfigure}[t]{\threecolwidth}
    \centering
    \includegraphics[width=1\linewidth, height=6.5cm, keepaspectratio]{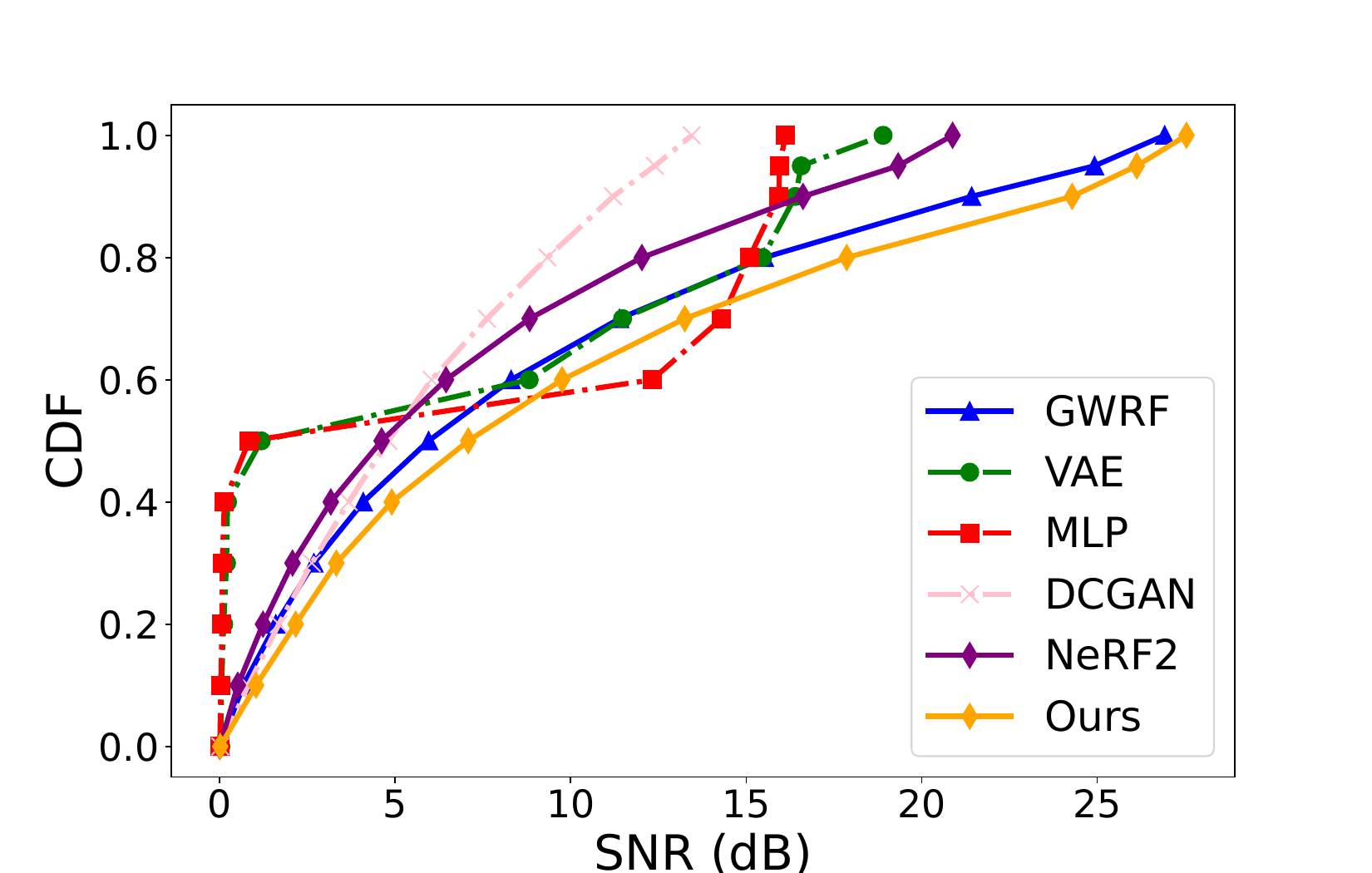}
    \caption{Conference dataset.}
    \label{fig:sub3}
  \end{subfigure}
  \vspace{0.5em}
  \caption{CDF-SNR performance comparisons across three CSI datasets.}
  \label{fig:three-wide}
\end{figure*}

Table~\ref{tab:performance_comparison} presents a comprehensive comparison between our proposed GAI-NeRF model and several baseline methods across diverse evaluation scenarios. To improve efficiency, we remove the self-attention mechanism in the tokenizer while keeping the overall architecture unchanged. The results demonstrate the effectiveness of our approach for wireless channel modeling. In the Room~1 dataset, GAI-NeRF achieves MAE values of 3.17~dB at 2.4~GHz and 1.75~dB at 5.0~GHz, marking substantial improvements compared with NeRF$^2$, which records 3.58~dB and 2.87~dB. Traditional deep learning methods exhibit weaker performance, with DCGAN yielding 4.01~dB and 3.39~dB, VAE reaching 5.75~dB and 5.49~dB, and the MLP baseline producing errors as high as 7.31~dB and 9.28~dB. The Room~2 dataset reveals a similar trend. GAI-NeRF achieves the lowest error rates of 3.06~dB at 2.4~GHz and 1.80~dB at 5.0~GHz, while NeRF$^2$ follows with 3.92~dB and 2.07~dB, and the gap further widens against DCGAN, VAE, and MLP. The performance evaluations across three specialized datasets provide the validation of GAI-NeRF’s robustness and generalization. 

In Fig.~\ref{fig:three-wide}, within the MIMO dataset GAI-NeRF reaches an SNR of 25.78~dB, very close to the strongest competitor GWRF at 25.08~dB, yet higher than NeRF$^2$ at 23.85~dB, VAE at 21.47~dB, MLP at 19.86~dB, and DCGAN at 13.12~dB. In the Bedroom dataset, where multipath propagation is especially complex, GAI-NeRF shows its most striking advantage, achieving 29.13~dB and outperforming GWRF at 28.35~dB while far exceeding NeRF$^2$, MLP, VAE, and DCGAN. A consistent pattern emerges in the Conference room dataset, where GAI-NeRF achieves 27.54~dB compared with 26.92~dB for GWRF, while the remaining baselines remain several decibels lower. These results highlight that GAI-NeRF not only achieves state-of-the-art accuracy but also delivers robust and physically consistent predictions across a wide range of wireless environments and frequency bands, underscoring the value of integrating geometric structure with electromagnetic interaction patterns in a unified learning framework.

\begin{figure}[htbp]
\centering
\includegraphics[width=0.8\linewidth]{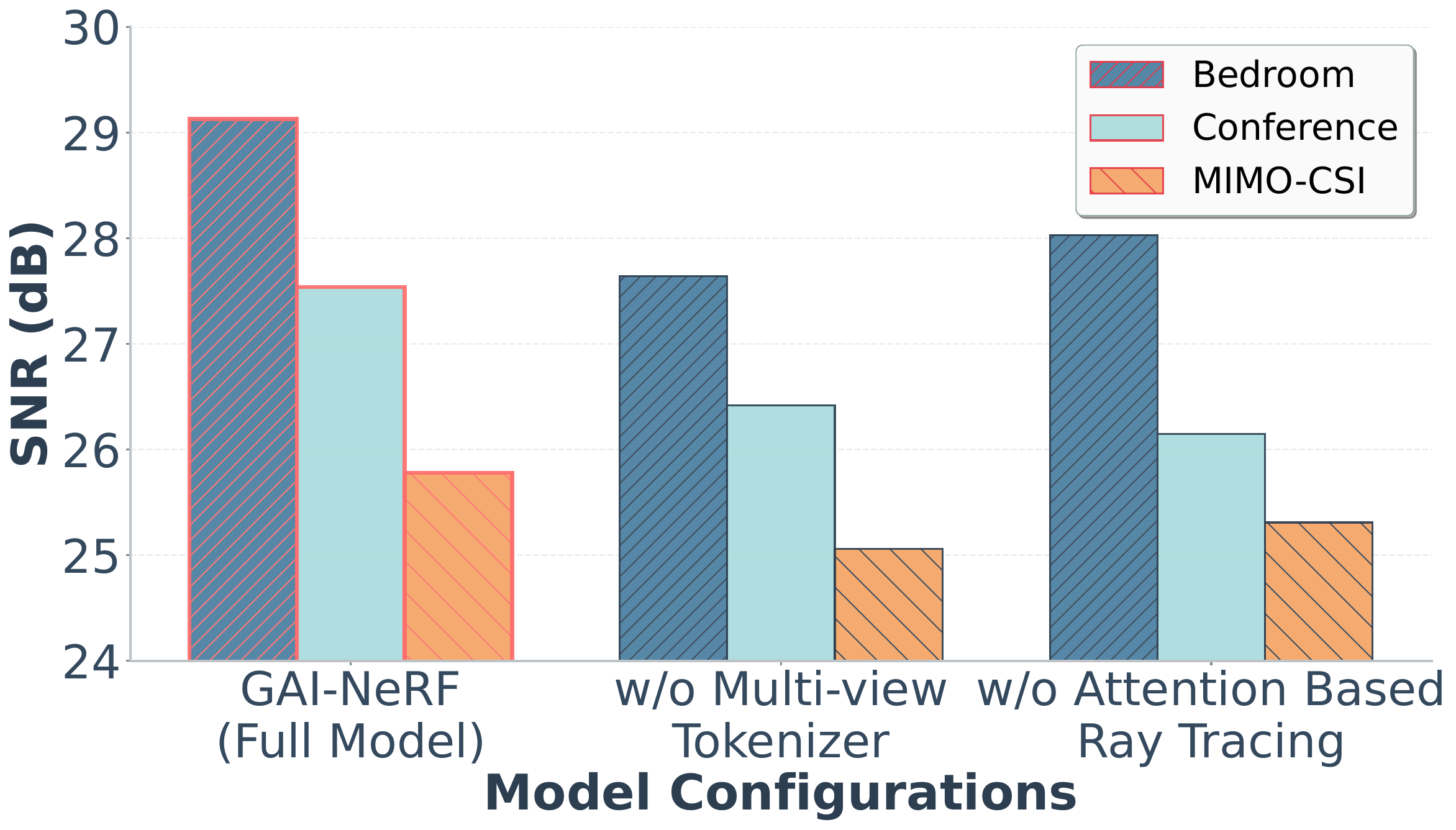} 
\caption{Ablation study on GAI-NeRF in CSI datasets.}
\label{tab:abla}
\end{figure}

\subsubsection{Ablation Study for CSI datasets}
Fig.~\ref{tab:abla} presents our ablation study results across three datasets: Bedroom, Conference, and MIMO-CSI. Our complete GAI-NeRF model achieves the highest SNR values of 29.13 dB, 27.54 dB, and 25.78 dB respectively. When removing the multi-view tokenizer component, performance decreases by 1.49 dB on Bedroom, 1.12 dB on Conference, and 0.72 dB on MIMO-CSI, demonstrating its importance for capturing spatial relationships. Similarly, excluding the attention-based ray tracing module results in SNR reductions of 1.10 dB, 1.39 dB, and 0.47 dB across the three datasets. These results confirm that both components contribute significantly to the model's performance, with the multi-view tokenizer showing slightly greater impact on the Bedroom dataset while the attention-based ray tracing module proves more crucial for the Conference environment.

\subsubsection{Generalization}

\begin{figure}[htbp]
\centering
\includegraphics[width=0.8\linewidth]{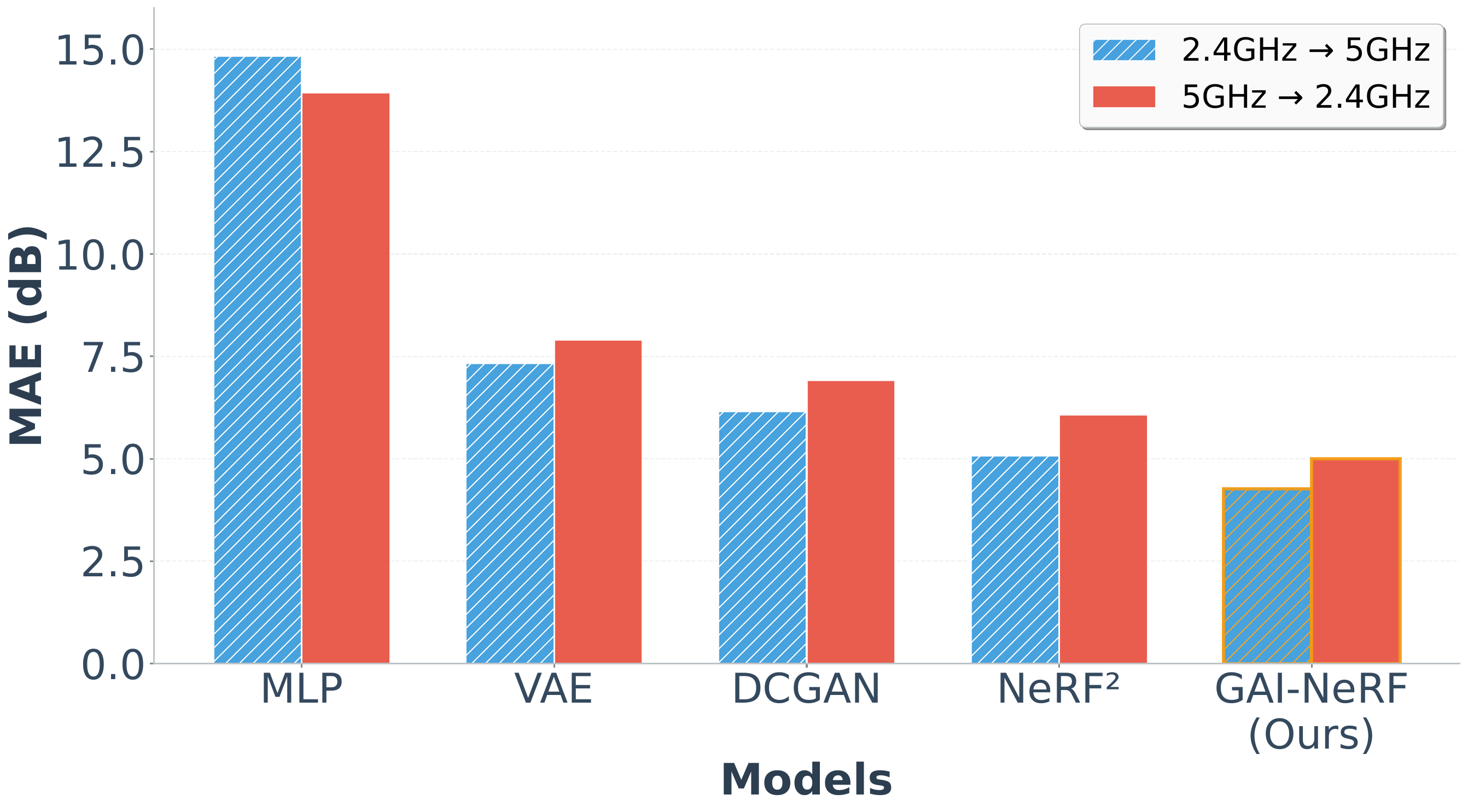}
\caption{Generalization performance comparison across
different models and frequencies in our own-built dataset. Results are averaged across both rooms (Evaluation metrics: MAE).}
\label{tab:model_performance}
\end{figure}
In this paper, we consider two generalization settings: frequencies and environment changes. Fig.~\ref{tab:model_performance} presents the generalization performance of our proposed GAI-NeRF model compared to baseline methods for cross-frequency channel prediction. We evaluate the MAE for bidirectional frequency translation between 2.4 GHz and 5 GHz bands. Results are presented with mean values in both rooms. GAI-NeRF achieves the lowest error rates in both directions, with 4.27 MAE for 2.4 GHz to 5 GHz prediction and 5.00 MAE for the reverse direction. This represents a significant improvement over the NeRF$^2$, reducing error by 15.8\% and 17.6\% respectively. The generative models (VAE and DCGAN) demonstrate intermediate performance, while the standard MLP exhibits the highest error rates, confirming the limitations of direct mapping approaches for cross-frequency channel prediction. These results validate that our integration of geometric awareness and implicit neural representations enables more accurate frequency translation compared to existing methods.

\begin{figure}[htbp]
    \centering
    \begin{subfigure}[b]{0.24\linewidth}
        \centering
        \includegraphics[width=\linewidth]{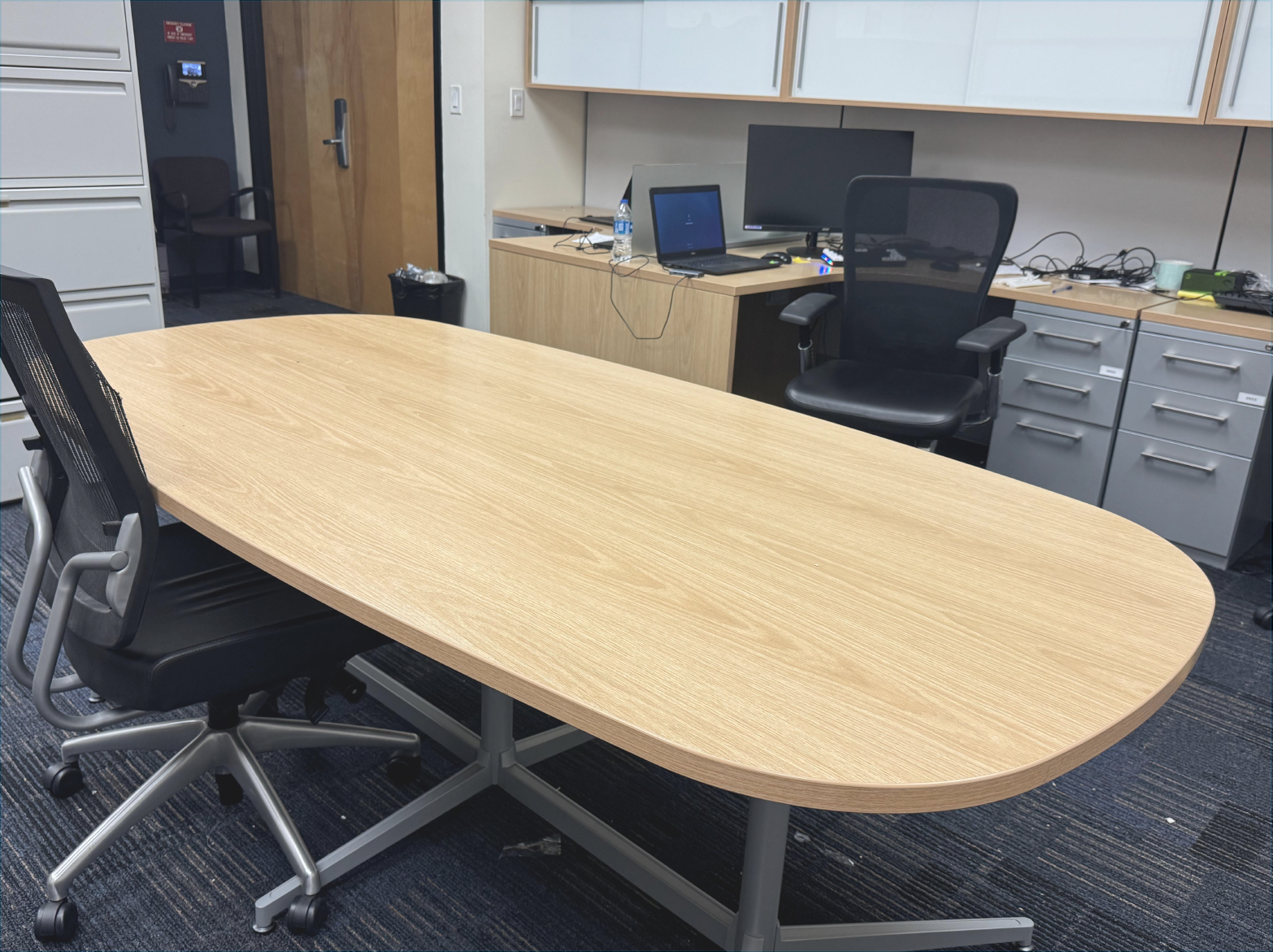}
        \caption{Initial Scene.}
        \label{fig:init_scene}
    \end{subfigure}
    \hfill
    \begin{subfigure}[b]{0.23\linewidth}
        \centering
        \includegraphics[width=\linewidth]{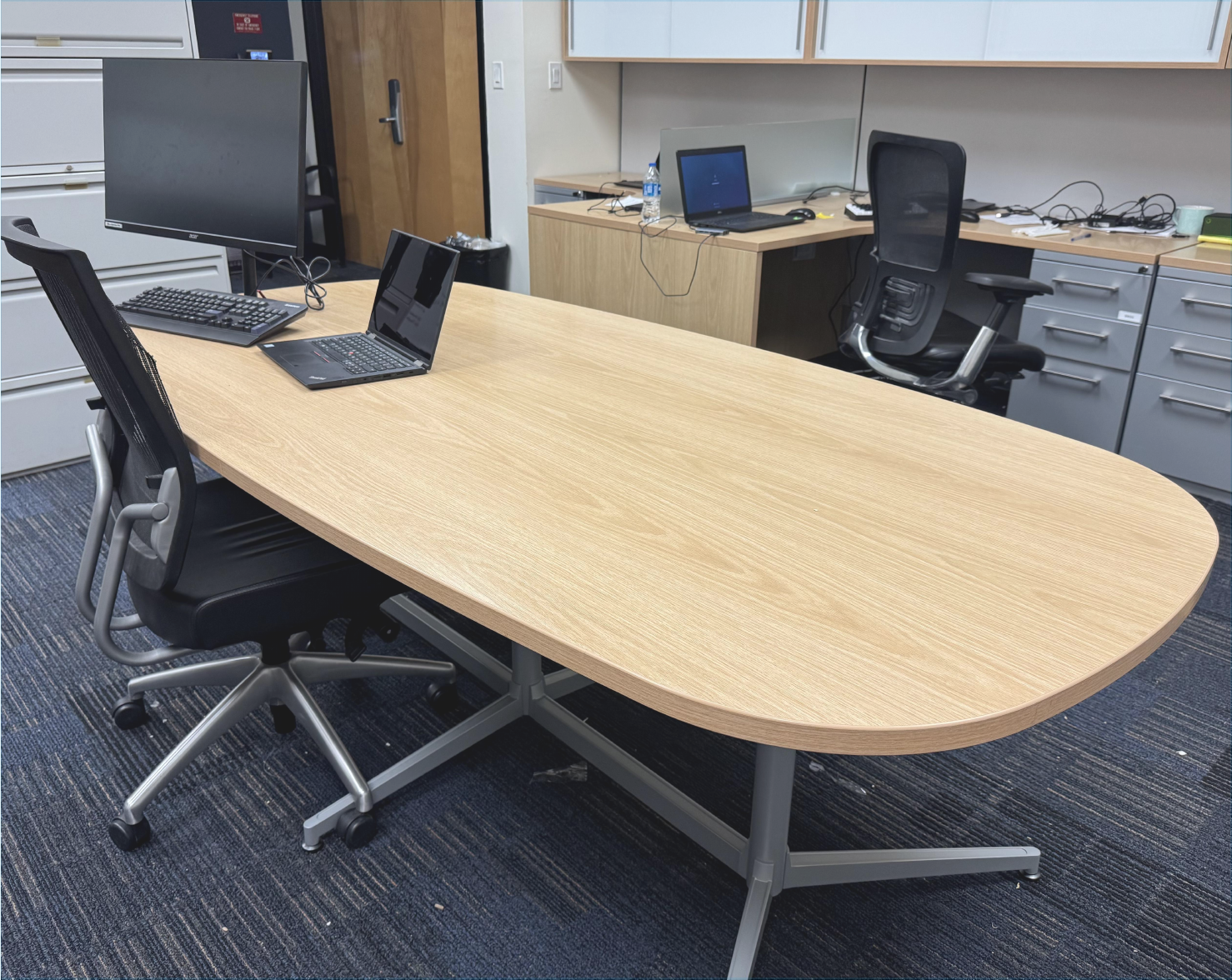}
        \caption{Addition.}
        \label{fig:scene1}
    \end{subfigure}
    \hfill
    \begin{subfigure}[b]{0.24\linewidth}
        \centering
        \includegraphics[width=\linewidth]{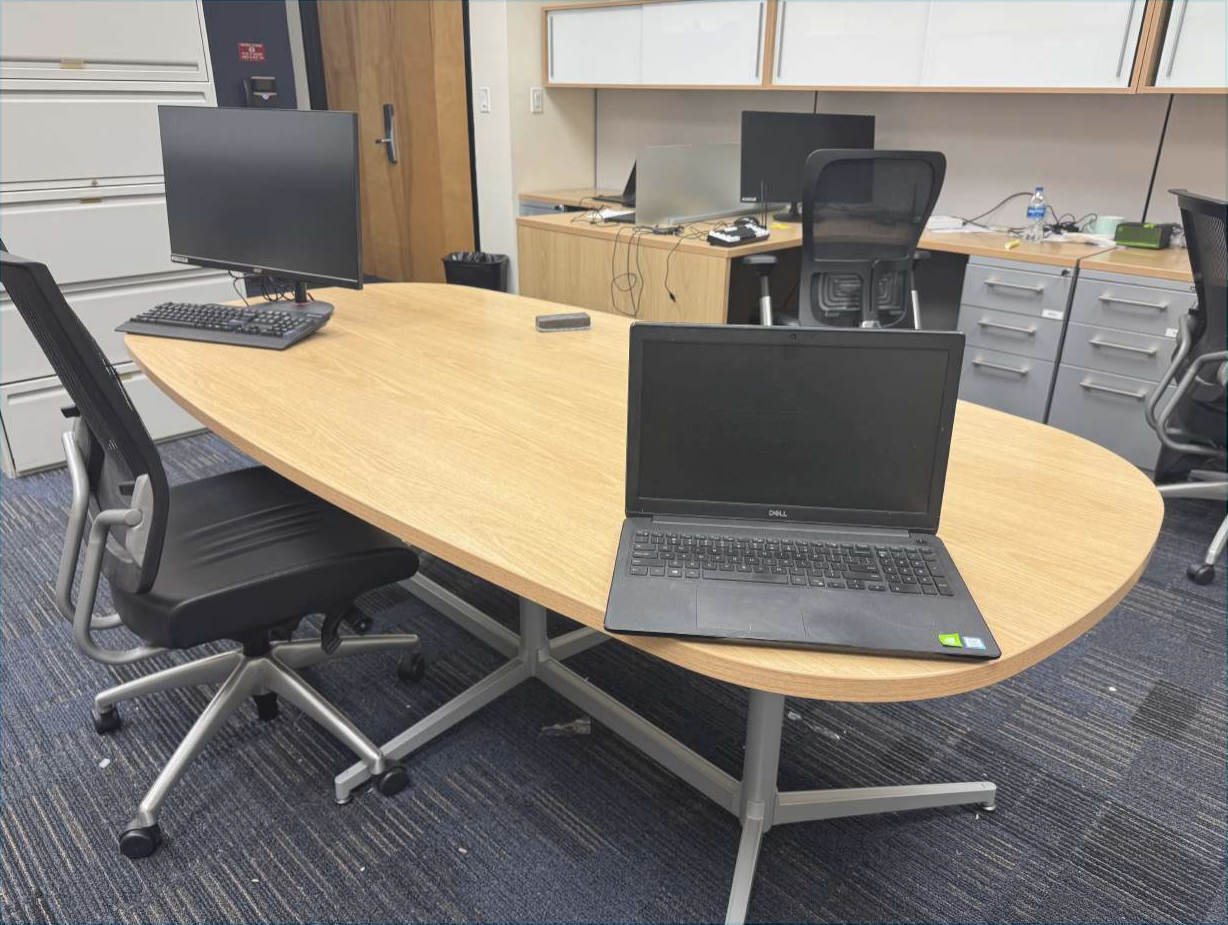}
        \caption{Relocation.}
        \label{fig:scene2}
    \end{subfigure}
    \hfill
    \begin{subfigure}[b]{0.24\linewidth}
        \centering
        \includegraphics[width=\linewidth]{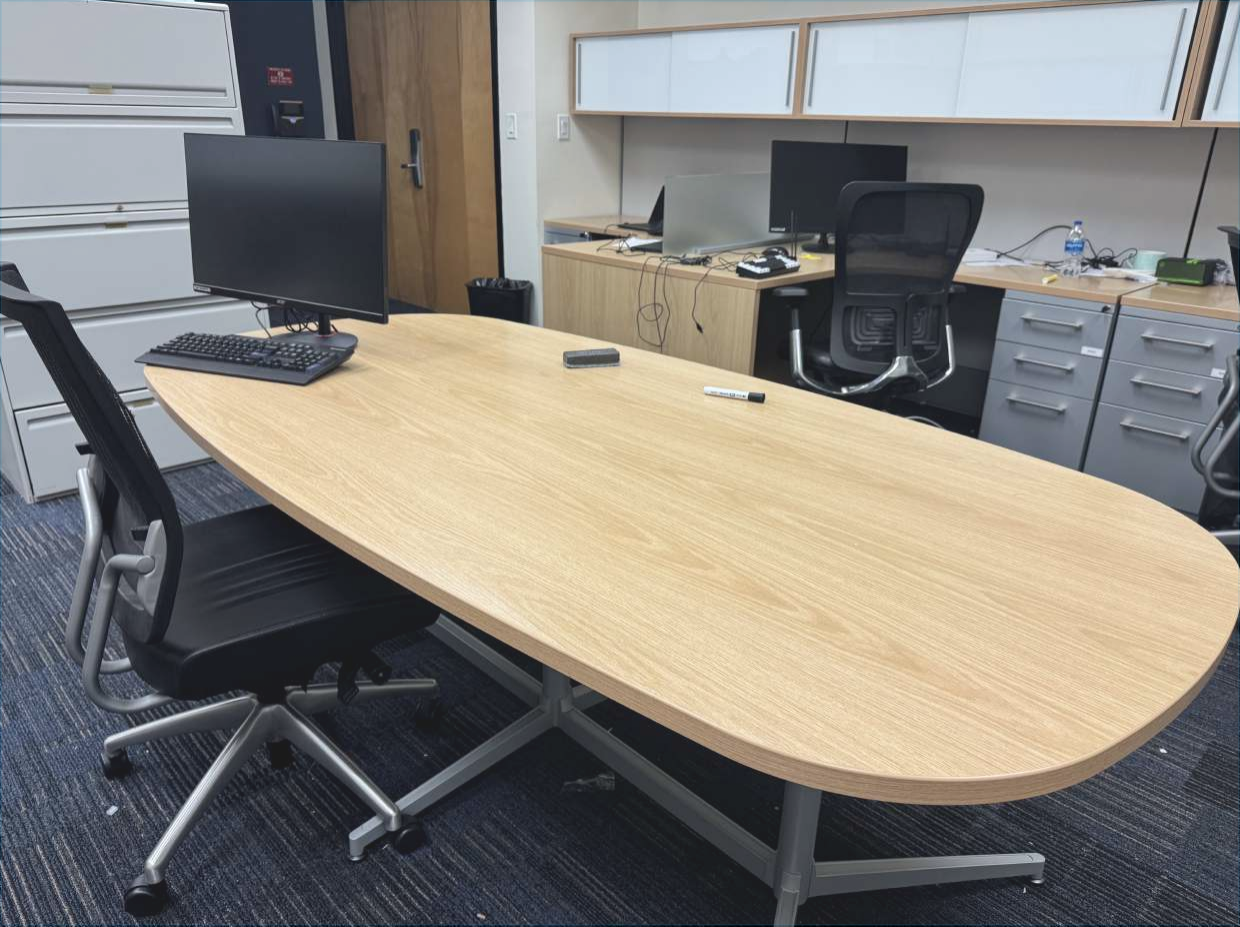}
        \caption{Removal.}
        \label{fig:scene3}
    \end{subfigure}
    \caption{Three generalization scenes are evaluated to test the framework's adaptability. Scene 1 involves adding a tablet set to the empty table. Scene 2 examines the impact of relocating the laptop to different locations. Scene 3 tests the removal of the laptop from the tablet setup. These scenarios assess the model's robustness under various environmental changes.}
    \label{fig:four-gen}
\end{figure}


Fig.~\ref{fig:four-gen} presents the four fundamental scenarios used to evaluate our model's generalization capabilities: the initial scene, addition of tablet devices, relocation of the tablet set, and removal of the laptop from the tablet configuration. Table~\ref{tab:generalization-performance} evaluates the generalization capability of our proposed GAI-NeRF model against baseline methods across three distinct scenes using RSSI data at both 2.4~GHz and 5~GHz. Across all scenarios, GAI-NeRF consistently delivers the lowest MAE values, for instance reducing errors in Scene~1 to 3.25~dB at 2.4~GHz and 3.37~dB at 5~GHz, while achieving similarly strong results of 3.48~dB and 2.33~dB in Scene~2 and 3.45~dB and 2.27~dB in Scene~3. These outcomes correspond to notable margins over the strongest baseline, NeRF$^2$, with improvements approaching 1~dB in several cases, and substantially larger gains over conventional models. Compared with traditional approaches, GAI-NeRF consistently narrows prediction error by wide margins—exceeding MLP by nearly 9–11~dB and outperforming VAE and DCGAN by several decibels—demonstrating that the integration of geometric algebra principles with multi-view learning provides not only numerical advantages but also robust generalization across frequencies and diverse propagation environments.


\begin{table}[htbp]
\centering
\footnotesize  
\renewcommand{\arraystretch}{1.3}
\setlength{\tabcolsep}{3pt} 
\caption{Detailed generalization performance (MAE in dB$\downarrow$) across different models and scenes on our RSSI dataset at 2.4 GHz and 5 GHz.}
\begin{tabular}{l@{\hskip 4pt}cc@{\hskip 4pt}cc@{\hskip 4pt}cc}
\toprule
\multirow{2}{*}{\textbf{Model}} & \multicolumn{2}{c}{\textbf{Scene 1}} & \multicolumn{2}{c}{\textbf{Scene 2}} & \multicolumn{2}{c}{\textbf{Scene 3}} \\
\cmidrule(lr){2-3} \cmidrule(lr){4-5} \cmidrule(lr){6-7}
& 2.4 GHz & 5 GHz & 2.4 GHz & 5 GHz & 2.4 GHz & 5 GHz \\
\midrule
MLP   & 11.92  & 9.65    & 12.52    & 10.08    & 13.51    & 10.21    \\
VAE   & 6.63    & 5.51    & 5.91    & 6.13    & 6.14    & 5.50   \\
DCGAN   &  4.61   & 4.48    & 4.91    & 4.33    & 4.23    & 4.17    \\
NeRF$^2$ & 4.34  & 3.79  & 4.14  & 2.93  & 3.67    & 2.97    \\
\textbf{GAI-NeRF (Ours)}  & \textbf{3.25} & \textbf{3.37} & \textbf{3.48} &
\textbf{2.33} & \textbf{3.45} & \textbf{2.27}    \\
\bottomrule
\end{tabular}
\label{tab:generalization-performance}
\end{table}

\section{Conclusion}
In this paper, we propose the first multi-view framework for wireless channel modeling, grounded in geometric algebra and wireless ray transmission theory to guide the learning of ray-object interactions. Our study demonstrates that integrating geometric scene structure with electromagnetic interaction patterns enables physically consistent channel predictions that more faithfully reflect the underlying propagation mechanisms. The attention-augmented ray tracing module proves essential for improving generalization, allowing the model to remain robust in dynamic environments. Beyond achieving state-of-the-art accuracy across both our newly collected datasets and existing benchmarks, our findings highlight a key insight: multi-view learning provides a principled pathway to unify geometric understanding and electromagnetic modeling within a single framework. This not only enhances predictive performance but also strengthens the theoretical foundation of wireless channel modeling.

\section{Acknowledgment}
This work is supported in part by the NSF (CNS-2415209, CNS-2317190, CNS-2107190, CNS-2415208, IIS-2306791, and CNS-2319343).

\bibliographystyle{IEEEtran}
\bibliography{ref}

\end{document}